\documentclass[a4paper,8pt,twocolumn]{article} 

\usepackage{graphicx}
\usepackage{color}
\usepackage{longtable}
\usepackage{url}

\newlength{\hfwidth}
\newlength{\hfwidthsingle}
\newcommand{\sect}[1]{Sect.~\ref{#1}}

\begin{document}

\title{\bf Naming the extrasolar planets}
\author{W. Lyra\vspace{.2cm}\\
{\it Max Planck Institute for Astronomy, 
     K\"onigstuhl 17, 69177, 
     Heidelberg, Germany}\vspace{.2cm}\\
wlyra@mpia.de}
\date{}
\maketitle

\begin{abstract}
  Extrasolar planets are not named and are referred to only 
  by their assigned scientific designation. The reason given 
  by the IAU to not name the planets is that it is considered 
  impractical as planets are expected to be common. I advance 
  some reasons as to why this logic is flawed, and suggest names for 
  the 403 extrasolar planet candidates known as of Oct 2009. The names
  follow a scheme of association with the constellation that 
  the host star pertains to, and therefore are mostly drawn from 
  Roman-Greek mythology. Other mythologies may also be used 
  given that a suitable association is established. 
\end{abstract}

\section{Introduction}

Since the discovery of the first extrasolar planet, around the star 
51 Pegasi (Mayor \& Queloz 1995), over 400 planets surrounding other 
stars have been discovered. It is no exaggeration to say that, for 
astronomy, the year of 1995 has a historic resonance with 1781, 1846, 
and 1930. However, unlike Uranus, Neptune, and Pluto, the almost 
totality of these extrasolar planets are known by no other name than the 
scientifically dry designations given to them. 

It is my intent to make the case that naming these planets is desirable. 
Poincar\'e (1905) emphasized the usefulness 
of astronomy by saying that ``it is useful because it raises us above 
ourselves, because it is great, because it is beautiful''. Planet 
MOA-2008-BLG-310-L b, a sub-Saturnian mass planet recently detected 
in the Galactic Bulge with the technique of microlensing 
(Janczak et al. 2009), 
certainly inspires this feeling of transcendence Poincar\'e describes. But 
its name hardly helps on conveying it.

One of the main reasons I consider for naming the extrasolar 
planets is the Copernican 
Principle itself. Our place in the cosmos is not special in any way, so 
there is no reason why only the planetary objects in the solar system 
should be named. Shakespeare would perhaps disagree with me and say that 
Io by any other name would smell as bad; and it is true that HD 128311 b 
will have the same radial velocity curve irrespectively of us naming it after 
a catalog number or after Bacchus. However, the non-special nature 
of our place in the Universe is better 
underscored by naming our neighbors. Mercury - Venus - Earth - Mars 
is a sequence of equals. Sol b - Sol c - Earth - Sol d would implicitly 
imply that the Earth is special in some way. Likewise, Jupiter is 
being paired to obscure names such as XO-1 b, TrES-4 b, and 
OGLE-TR-182 b, which does not help educators 
convey the message that these planets are quite similar to Jupiter. 
In stark contrast, the sentence``planet Apollo is a gas giant like Jupiter'' 
is heavily - yet invisibly - coated with Copernicanism.

One reason given by the IAU for not considering naming the extrasolar planets 
is that it is a task deemed impractical. One source is quoted as having 
said ``if planets are found to occur very frequently in the Universe, a 
system of individual names for planets might well rapidly be found equally 
impracticable as it is for stars, as planet discoveries progress.''
{\footnote{\url{http://www.iau.org/public_press/themes/extrasolar_planets/}}}. 
This leads to a second argument. 
It is indeed impractical to name all stars. But some stars are named 
nonetheless. In fact, all other classes of astronomical bodies are 
named. Galaxies are named. Stars are named. Even asteroids 
are named. So why not name the exoplanets? Granted, not all galaxies 
have names. Only the nearby ones, and in this case, the names are 
easily fixed due to the shape of the galaxy - Whirlpool, Antenna, 
Sombrero -, or of the constellation - Andromeda, Circinus, Carina. 
Star naming also has a criterion - the brightness. All stars 
brighter than $V$=1.5 have a proper name, the frequency of named 
stars declining as the magnitude increases{\footnote{S. Boscardin, 
from Observat\'orio Nacional, Rio de Janeiro, Brazil, pointed to me 
that there are four unnamed stars brighter than $V$=2.5. They are gamma 
and delta Velorum, of magnitudes $V$=1.8 and 2.0, and epsilon and eta 
Centauri, both of magnitude 
$V$=2.5. I credit him here for this information, and echo his 
suggestion that the IAU should consider naming these four lone stars. He 
further informs me that there are 84 stars between 2.5$<$$V$$<$3.0, 
51 of which are named; and 106 stars between 3.0$<$$V$$<$3.5, 33 of 
which are named.}}. 
Yet nothing has stopped people from naming over 15,000 
asteroids and minor planets. In fact, it seems to be that 
the main reason for over 400 known 
exoplanets remaining unnamed is that no one has yet done the 
job of naming them all. Indeed, as discoveries proceed (and hopefully 
skyrocket with the Gaia mission), naming all planets will be impractical. 
But the benefits of having 
some of them named is as clear as in the case of stars. In this 
manuscript, I set myself to the task. 

In some cases, planets have already been nicknamed. For some, the 
epithet is sound. On naming 51 Pegasi b, I immediately thought 
of Bellerophon, the rider of the winged horse. Then I found out that 
someone else also shared the same taste for mythological associations 
and had already nicknamed the planet with the same name. I also came 
across the webpage of the ``Extrasolar planet naming society'', created 
in November 2008 with the idea of organizing a concerted collective 
effort to name the known exoplanets. Unfortunately, it did not seem 
to have gained any momentum. Individual efforts seem to beat collective 
ones in this case. I highlight here the webpage of Devon Moore, where  
over 40 exoplanet names have been suggested 
{\footnote{\url{http://nuclearvacuum.wikia.com/wiki/Names_for_extrasolar_planets}}}. 
Some of the names Moore chose are indeed good, some I have reasons 
to object. I will go back to that later. 

Discoverers have also attempted to name the fruit of their labor. 
The three planets around Upsilon Andromedae have been nicknamed 
``Fourpiter'', ``Twopiter'', and ``Dinky''. These names were suggested by a 
class of 4th graders, and have the advantage of carrying information on the 
planets' size.  However, they are tantalizingly unpalatable to those of a more 
classical mind. No offense to the very creative children, but apparently 
long gone are the times of Venitia Burney, then 11 years old, who suggested 
the name of Pluto, the Roman-Greek god of the underworld (not the Disney dog), 
for the distant, cold, (ex-)planet discovered in 1930. 70 Virginis b was 
similarly nicknamed ``Goldilocks'', for its position in 
the habitable zone of its star. One could imagine that had the discovery 
been made by a Swedish team if they would have called the planet 
{\it Lagom}, after their unique word for ``just right''. Not to mention 
the internal names used by discovery teams that sometimes leak to the media, 
such as ``Xena'', ``Easterbunny'' and ``Santa'' (the Kuiper Belt 
objects Eris, Makemake, and Haumea, respectively). 

In this manuscript, I advocate a return to classic tradition, and 
propose a simple way to name the known 
exoplanets, exposed in Sect 2. It basically consists of giving them names from Roman-Greek mythology
associated with the constellation that the host star pertains. 
The columns in Table~1 show, respectively, the names of the planets 
listed by constellation, the name used in the astronomical literature, 
their masses in units of Jupiter's mass ($M_J$), semi-major 
axes in astronomical units (AU), eccentricities, right ascension, declination, and, finally, the 
proposed name{\footnote{Table~1 can also be found online at \\\url{http://www.mpia.de/homes/lyra/planet_naming.html}}}. The mythological associations for these names are 
explained in Sect 3. 

The method used was the following. I got the planets from Jean Schneider's
{\it The Extrasolar Planets Encyclopaedia}, and listed the planets by constellation. I then suggest names for the planets with the help of a dictionary of 
Greek mythology (Dixon-Kennedy 1998) and extensive use of Wikipedia. 
It is commonplace that information in Wikipedia is not always trustable, and 
has to be used with care. That is correct, but when well-handled, the 
information there available is vital. For instance, listing the planets 
by constellation would have been quite time-consuming without Wikipedia, since 
the papers not always mention it, giving only the coordinates. To my 
surprise and amazement, almost all extrasolar planets have a wiki, that 
also informs in which constellation the star is located. I also made use 
of the following online source, \url{http://www.theoi.com/}, that works 
as an online interactive dictionary of Greek mythology. 

\section{The naming convention}
\label{sect:naming}

In assigning the names, it seems natural to follow the mythological 
stories of the constellations. Andromeda's 
myth for the planets in Andromeda, Hercules' myth for planets 
in Hercules. It is as simple as it sounds, though there are 
some caveats where subtleties apply.

In suggesting names for the planets, I was tempted to 
circumvent some problems that, in the end, proved unavoidable. One 
of them refers to too common personal names. For instance, 
Leda is a natural choice for a planet in Cygnus, the Swan. 
The name is not popular in the United States, but is quite 
popular in Brazil. I also assume that most native English 
speakers would not favor a planet called ``Jason'' even 
though Carina, Puppis, and Vela beg for the name.

However, any attempt to avoid these cases will invariably 
be doomed to failure. When I finished an earlier draft of this 
manuscript, I asked a Greek colleague to go through Table 1 and 
check that the names were not too common in modern Greece. She
pointed a handful of semi-common names, a singer's last name, and added 
that Peristera (\sect{Columba}) simply meant 
``female pigeon'', and that it sounded ``a little funny''. The problem, 
of course, is that one can only identify these special cases in 
languages and cultures one is familiar with.

The same problem exists for cacophonies. Take Uranus, for instance. 
It is an endless source of humor for each generation of ten-year old 
native English speakers, yet innocuous in any other language.
This is a well known problem for commercial companies, since their 
well-tailored name may well be a common name or sound offensive 
in another language. Avoiding it in all languages would be virtually 
impossible. As one cannot guarantee that no single name picked is a 
common name or offensive word in any other language, it is less subjective 
to not avoid it even when one can identify it. Speaking of companies, 
I also made no attempt to avoid overly used commercial names such 
as Clio (\sect{Reticulum}), Python (\sect{Serpens}), or 
Nike{\footnote{Though I add that Nike 
should be read {\it nikki}, not {\it nyke}. Obviously the same 
applies to Dike, (\sect{Libra}), here stated lest I start 
getting emails of protest from conservative 
sects of society.}} (\sect{Pegasus}). 

I did, however, perform some censorship. For instance, 
Electryon (one of the Perseids) sounds too much like 
electron (ispell even tried to correct it). Certainly we do 
not want to have high-school teenagers confused over if electron 
is a sub-atomic particle or a planet. Aesthetics also played 
a role, since many times when faced with two or more choices, I 
picked the names that I judged more beautiful. I imagine that an 
observational proposal to image planet Halirrhothius can be avoided. 
And planet Medusa would be just short of petrifying some. I am aware 
that this is subjective, and other persons would have different 
aesthetic sensitivities. However, I could not avoid it and will 
not pretend I did. Though I notice that 
when reading the names over and over again, I tend to get used to 
them. In fact, I am already starting to like the sound of 
Halirrhothius. I invite the reader to do the same exercise with 
unappetizing names that eventually show up in Table~1.  

Another problem is that most names are already used for asteroids. 
Repetition in 
this case is unavoidable, since, as mentioned before, over 15,000 of them 
are named, which exhausted most of the 
mythological names. The policy is certainly not ideal. Yet, it is 
far from unacceptable. The ambiguity in names is actually already 
in vogue. Atlas is both a star of the Pleiades and a satellite 
of Saturn. We may doubly honor a deity in the same way that we 
may use the same name for cities in different countries. The post 
office is hardly ever confused between Memphis, Tennessee, 
and Memphis, Egypt. We also have Paris in Texas, Moscow in Idaho, 
Dublin in Ohio, London in Ontario, San Jose in California, Panama 
City in Florida, Birmingham in Alabama, York in Pennsylvania, 
Manchester in New Hampshire, Cambridge in Massachusetts, 
and Rome in New York, just to name a few. The point is that 
the fields are far enough apart that the two homonymous celestial bodies 
will rarely be mentioned 
in the same sentence. And should the unlikely case arise, I see little 
trouble in using a couple more words to clearly state the difference, 
i.e., ``Leda, giant planet of HAT-P-7'', and ``Leda, minor satellite of 
Jupiter''. However, I do refrain from using names of the major moons. 
So it is that Triton, Titan, and the Galilean moons shall not be used. 

The only point where I deviate from the norm of associating the 
planet name with the myth of the constellation where the host star 
pertains is in the case of the pulsar planets. In the paucity of visible 
light, these planets are sunk in darkness, and I suggest they be named 
after the souls of the damned that inhabit the Tartarus. So it 
is that I suggest the planets of the pulsar PSR 1257+12 should be named 
after Sisyphus, Ixion, and Tantalus. The planet of PSR B1620-26, thought 
to be extremely old, may be named after Erebus itself, 
since it is a primordial deity that represents darkness. 

\subsection{Comparison between this and Moore's suggestion}

In Moore's webpage, as mentioned, $\sim$50 planet names were 
already suggested. I acknowledge Moore for anticipating in some way the 
method that I propose here, i.e., naming the planets after some 
association, even loose, to the constellation 
of the host star. So it is that the quadruple system around Gliese 581, 
in the constellation of Libra, may as well be named after Themis, the 
Titanian goddess of justice, and her offspring, the Horae. Eirene (peace),
Dike (trial), and Eunomia (rule of law). Moore had already suggested Dike 
for the planet 23 Librae b, but to have Themis' offspring following her 
seems a more natural choice. Another point of coincidental agreement 
regards the already mentioned planets around Upsilon Andromedae. Both 
Moore and I independently found it natural to name them after the descendants 
of Andromeda. We only chose different sons. While he chooses Alcaeus and 
the confusing name of Electryon, I prefer Heleus, Mestor, and Cynurus. 
It is my sincere hope that the discoverers find them better than 
Fourpiter, Twopiter, and Dinky. 

A point that I disagree, however, is the policy of reserving a good 
name for yet-to-be-discovered planets. Moore does so with Orpheus, 
reserving the name for a planet around Vega. I am not insensitive 
to the poetry of the choice. Not only Vega is the brightest star of 
Orpheus' instrument, Lyra, the lyre, but it will also soon leave the 
main sequence. Vega's death would sink planet Orpheus into 
darkness, thus astronomically mimicking his famous descent 
into the underworld. Yet, no matter how likely, we cannot 
pre-assume that a planet will be found in Vega. I instead use 
Orpheus to name the planet HD 173416 b, also in Lyra. Below I 
discuss another main point of disagreement.

\subsection{Non Roman-Greek mythologies}
\label{subsect:nonroman}

Many of the names suggested by Moore come from other mythologies rather
than Roman-Greek. In particular, some of the planets were dubbed 
Yahweh, Jehovah, and Satan; perhaps following the ``Metuselah'' 
suggestion for the planet PSR B1620-26 b. I argue that names from 
Judeo-Christian tradition are to be avoided, due to their still widespread 
acceptance through the world in the current age. Adherents of the 
Abrahamic religions - and they go by the billions - certainly will 
not appreciate a planet named after Lucifer, and most likely will not like 
to see the name of their supreme deity being given to a ball of 
rotating gas. 

The discussion about keeping or not keeping the classic tradition 
is by no means new. Galileo named the satellites he discovered 
{\it Medicia Sidera}, after his patron. Cassini followed his exampled 
and named the satellites of Saturn that he discovered {\it Sidera Lodoicea} 
(Louisian Stars), after king Louis XIV of France, also his patron. 
The naming of Uranus and Neptune is a fascinating chapter in the 
history of Astronomy, and to that I refer the reader to the excellent essays
of Gingerish (1958) and Kollerstrom (2009). It suffices to say 
that the former was known as ``George's star'', 
``George's planet'' or ``Herschell''  well into the mid-19th century, and 
the latter went by the name ``Le Verrier'' even though Le Verrier himself 
suggested the name by which we call it today.

The main argument given at the time for not keeping the 
mythological tradition was based on the fact that the classical 
planets (i.e., up to Saturn) were named by the ancients after their 
gods in an era long gone. William Herschel is quoted as saying that 
``In the present more philosophical era it would hardly be allowable 
to have recourse to the same method'' (Dreyer 1912). Bode, however, 
argued for continuity, and remarked that just as Saturn was the father 
of Jupiter, the new planet should be named after the father of Saturn, 
Uranus. Uranus was also the consort of Gaia, the Earth. Since Venus sprang 
from Uranus, and both Mercury and Mars are Jupiter's children, the planets 
would coherently represent the mythological family. 

Of course, that debate 
revolved around whether or not Roman-Greek classical tradition should be
continued, whereas here the discussion is whether or not to include 
other mythologies rather than Roman-Greek. However, we can once again 
invoke Bode's argument. The method proposed here is to name the 
planets after some association to the constellation of the host star.
As the constellations defined by the IAU are based on 
the Roman-Greek myths, the use of non Roman-Greek mythological 
names would break the coherency of the system. 

There is one important exception though, occupying almost 
half of the sky. It is that of the southern constellations, most of 
which have no mythological connection. In these cases, if a suitable 
association can be found, names from other mythologies may be 
used without abandoning coherency. Moore also touches this point, 
for example, when suggesting a planet in Vulpecula, the fox, be called 
Anubis, after the jackal-headed Egyptian deity. He also suggests the 
name Qilin, after the Chinese mythical animal, for a planet in 
Monoceros, the unicorn. However, neither 
a jackal is a fox nor the qilin a unicorn, so I do not use these 
names. Nevertheless, loose associations with other mythologies are 
welcome and desirable, since they maximize the available names.

Further associations may be drawn from the fact that 
many Roman-Greek heroes, gods and demigods find counterparts 
in other mythologies. For instance, in Norse mythology, Odin 
plays the role of Jupiter, Tyr that of Mars, Hela the role 
of Pluto, Freya that of Venus, Njord that of Neptune, and 
Sigurd (Sigfried) that of Hercules. Sun-gods feature in most 
mythologies and may thus be easily associated whenever Helios 
is invoked. Also, the naming system is of associations between myth and 
constellation, so nothing impedes associations 
with the way other ancient cultures divided the sky. For instance, 
Orion is Gilgamesh in Sumerian mythology. Chinese and Indians also 
divided the sky differently, and names may be draw from Chinese or Vedic 
mythologies according to those constellations. Name suggestions 
from astronomers better acquainted with these mythologies are of course 
welcome.

However, even though I am sympathetic to the inclusion 
of Norse, Egyptian, and Sumerian mythologies (to name 
the most popular ones after Roman-Greek), as long as they do 
not break the symmetry of the system, attention should be draw to 
the fact that it is not obvious where to draw 
the line on which mythologies to include and which not to. In the 
case of the trans-Neptunian objects, names such as Makemake (from 
Easter Island mythology) and Haumea (from Inuit mythology) were 
adopted. But one particular insight that usually goes unstated 
is that the well-intended honoring may instead be embarrassing 
for cultures that, unlike the Greeks, are not used to the 
international spotlight. I do not know what the feelings of 
the inhabitants of Easter Island and Greenland are on this 
particular subject, but I suspect most Brazilians would object 
to a planet named after the Headless Mule.


\subsection{The southern constellations}
\label{sect:southern}

The classical constellations described in the Almagest (Ptolomey 148) were 
of course drawn over the stars visible from the mid-latitudes of the northern 
hemisphere only. For these, the myths where to look for prospective 
names are obvious. However, most of the austral constellations defined 
during the Age of Navigation have no mythological association. I acknowledge 
Moore for also touching the important point of what to do with them. 

The step here is to find an association between them and classical 
mythology. An obvious one is Horologium, the hourglass. He suggested 
Cronus, Greek name of Saturn, Titan god of time, for the planet of 
Iota Horologii. I agree with the suggestion. These constellations 
deserve further discourse. 

\subsubsection{Plancius' constellations}
\label{sect:plancius}

Petrus Plancius, Dutch cartographer and uranographer, 
mapped the stars of the southern 
sky into new constellations. He asked a navigator to draw the positions of 
the stars around the south pole, and arranged these stars as he saw fit. 
Those were later published by Johann Bayer 
in his sky atlas, the Uranometria (Bayer 1603). Of these 
constellations, twelve have become standard. 

The drawback is that Plancius referred these constellations not to mythology, but to 
natural history. To that he added Indus, the Indian. Both follow the 
enthusiasm of the time over the newly discovered lands. His constellations 
are Apus, the bird of paradise; Chamaeleon; Dorado, the goldfish; 
Grus, the crane; Hydrus, the small water snake; Indus, the Indian; Musca, 
the Fly; Pavo, the Peacock; Phoenix; Triangulum Australe, 
the southern triangle; Tucana, the toucan; and Volans, the flying fish.
He later also defined Columba, the dove. 

These constellations are easily associated with Roman-Greek mythology. 
The sea animals may draw from the myth of Neptune. The land animals 
from Diana or Mercury. Diana for her role as protector of wildlife (in 
spite of being a hunter herself), Mercury for this role as protector of 
shepherds and flock animals (in spite of being a trickster himself). 
Grus (\sect{Grus}), Pavo (\sect{Pavo}), Phoenix (\sect{Phoenix}), and Columba 
(\sect{Columba}) 
have other interesting associations on their own, 
namely with the Deluge, Juno, and Helios, respectively. Triangulum 
(\sect{Triangulum}), Indus (\sect{Indus}), and Hydrus (\sect{Hydrus}) also 
draw from other associations. 

\subsubsection{Lacaille's constellations}
\label{sect:lacaille}

Nicolas Lacaille, french astronomer, performed observations of the southern 
sky from South Africa, from where he catalogued many stars and defined 
new constellations (Lacaille 1763). Fourteen of these are still with us today. 
These are namely Antlia, 
the pump; Caelum, the chisel; Circinus, 
the drawing compass; Fornax, the furnace; Horologium, the hourglass; Mensa, 
the table; Microscopium; Norma, 
the square; Octans, after the astronomical octant; Pictor, for the 
painter's easel; Pyxis, the mariner compass; Reticulum, Sculptor, and Telescopium. 

As can be seen, Lacaille followed Plancius in discontinuing 
the mythological tradition and named these faint regions of the 
sky mainly after scientific inventions of the time (Mensa is 
named after the Table Mountains, in South Africa, that impressed him. 
Lacaille is further credited to splitting the gigantic constellation 
of Argo Navis into Carina, Puppis, and Vela).

I propose here to in some way revert this action and name planets in  
these non-mythological, engineer-related, constellations, after 
inventors in the Greek myths. The first among them is of course 
Vulcan, the Olympian smith-god. Other god is Mercury, 
who invented the 
lyre, coinage, weights and measures, also sometimes related to 
sciences and inventions. Minerva, of course, goddess of 
wisdom, reverenced in most universities. Among mortals, Daedalus, 
father of Icarus, who constructed the labyrinth to trap the Minotaur, 
and also attempted flight. Yet another is Palamedes, who 
invented the dice during the siege of Troy according to Sophocles; 
Philoctetes, a master-archer, who invented other types of bow.

\subsubsection{Hevelius' constellations}
\label{sect:hevelius}

Johannes Hevelius defined ten new constellations (Hevelius 1690), seven of 
which are still in use today. Two of them are Canis Venatici, the hunting 
hounds, that he separated from Bootes; and Leo Minor, that he separated 
from Leo and defined as a lion cub that accompanies it. The other five 
of the new ones are Lacerta, the lizard; Lynx; Scutum, the shield; Sextans; 
and Vulpecula, the fox. 

Canes Venatici and Leo Minor have their mythological associations bound to 
the constellations whence they sprung. He did not devise mythological 
associations for the other five, but we can establish 
them nonetheless. As animals, Lacerta, Lynx, and Vulpecula could be 
associated with Diana and Mercury, as suggested in \sect{sect:plancius}. 
However, Lynx and Lacerta better resonate with Ceres, who transformed Lyncus, 
the king of Scythia, into a lynx. In another myth, she transformed Abas, 
a prince of Eleusis, into a lizard. Vulpecula finds echo with the myth 
of the Teumessian fox. Scutum has no planets discovered there yet, but as 
a shield, could be linked to Vulcan. As for Sextans, its being an instrument
and thus fruit of human intelligence, allows for an association with 
Minerva. For its astronomical relevance, with Apollo. However, I prefer to 
use names from the epics of navigation, the Odyssey being the most famous. 

\subsection{Navigation Myths and the Age of Discoveries}
\label{sect:lusiad}

The sources of names here proposed for the planets naturally come from the 
classical references of Roman-Greek mythology. Theogony (Hesiod ca\,700\,BC),
the Iliad (Homer ca\,800\,BCa), the Odyssey (Homer ca\,800\,BCb), 
Argonautica (Apollonius ca\,250\,BC), 
the Aeneid (Virgil 19\,BC), and Metamorphoses (Ovid 8\,AD). 
The whole problem with the southern 
constellations is that they were drawn almost two thousand years later, 
during the Ages of Discoveries, when those stars were first seen by 
inhabitants of the Northern hemisphere. 

As such, I am inclined to consider other sailing myth as well, namely, 
the Lusiad (Cam\~oes 1572). This master-piece 
of Portuguese literature{\footnote{The fact that many Portuguese 
astronomers are formerly or currently associated with the Geneva group did not 
affect my choice to include the Lusiad. If it also serves the purpose 
of homage, so be it.}} by Luis de Cam\~oes (anglicized {\it Camoens}) 
tells the story 
of Vasco da Gama's sailing to India in much the same spirit of 
Homer's Odyssey and Virgil's Aeneid, i.e., as a tale of dispute between the 
Greek gods, and how they interfere on and control human affairs.
Pertaining to the Renaissance, it is not part of 
standard classical Roman-Greek mythology. Yet Thomas Bulfinch found it 
an epic of enough importance to mention it in his {\it Age of Fable} 
(Bulfinch 1855).

The long poem starts after Vasco da Gama had already set sail to uncharted 
waters. Jupiter recognizes that this represents the coming of a new age 
in human 
history, the Age of Discovery, and summons a council of the Olympians to 
decide if they should help da Gama on reaching his goal. Jupiter gives 
his consent, as he knows that the Portuguese are destined by the Fates 
to reach the Indies. However, Bacchus fears that his cult will lose 
power if da Gama succeeds, and disagrees with the decision of Jupiter.  
Venus favors da Gama, as she sees the Portuguese as successors of the 
Romans, and knows that they will worship her as fervently as the ancients. 
A heated discussion soon starts, with the other gods taking either Venus 
or Bacchus' side, until Mars aggressively takes the word. He also takes 
the side of the Portuguese (for love of Venus or proud of the courageous 
sailors), and reminds the others 
that Jupiter had already decided. Jupiter agrees, and the council is finished.
The gods will help da Gama in his way to the Indies, while Bacchus, the 
main antagonist, tries 
in infinite ways to stop him. The epic is full of mythological allegories, 
such as Venus sending her nymphs to seduce the winds, thus easing the voyage, 
a Gigante guarding the Cape of Good Hope, and a Homeric
invocation to the nymphs of the river Tagus in the very first page. It 
draws from classic tradition even in geometry: the climax, 
the arrival in India, was placed at the point in the poem that 
divides the work according to the golden ratio. 

Pertaining to the time when the southern constellations were drawn, 
it is well suited to the role of source of names to their planets, 
without invoking non-classical themes.  Using a text written in the 
Renaissance for a constellation defined in the Renaissance 
provides a pleasant satisfaction for anyone who admires symmetry. 

I am aware that by including the Lusiad I might be opening the door to other 
myths rather than classic ones, but I once again stress that I 
choose the Lusiad explicitly because it is unique in Renaissance
literature in its close association with the Greek myths. Other  
works (renaissantist or not) such as France's {\it The Song of Roland}{\footnote{Unknown author, ca.\,1150}}, 
Germany's {\it Song of the Nibelungs}{\footnote{Unknown author, ca.\,1200. 
Even though the Nibelungenlied is excluded due its lack of association 
with Roman-Greek mythology, it can later be included due to the presence 
of Norse mythology.}}, 
Italy's {\it Divine Comedy} (Alighieri 1321), Spain's {\it Don Quixote} 
(Cervantes 1605, 1615), Hungary's {\it Siege of Sziget} (Zr\'inyi 1651), or 
Finland's {\it The Kalevala} (L\"onnrot 1849), to name some, do not 
meet this requisite. 

There is one, though, that does. It is  
{\it The Columbiad}, written by Joel Barlow (1807). 
Written in the early 19th century, it aspired to be a national 
epic to the nascent USA. Like in the Lusiad, it tells a modern 
sailing voyage in classic style. The voyage is of course that of 
Columbus and the discovery of the New World. Also drawing from 
Roman-Greek mythology, Hesper(is), the Evening (thus 
West) appears in vision to Columbus. Inspired by that 
vison, he sails West. Naturally, Hesperis symbolizes the 
undiscovered lands of the American continent. Unlike the Lusiad, though,
it did not gain the same popularity among its countrymen. 
One source (Vickers 1998) is quoted as saying that Barlow's  
heroic couplets ``attempt to emulate the more skillful Alexander Pope''. 
His reception aside, the classical inspiration of the Columbiad 
suffices for the purpose of this essay. Furthermore, just as the Lusiad 
is an excellent association for Mensa (\sect{Mensa}), the 
Columbiad pairs neatly with the constellation of Indus (\sect{Indus}),
 the Indian, that represents a native of the Americas.

\subsection{Former constellations}
\label{sect:former}

In two cases, Camelopardalis (\sect{Camelopardalis}) and Microscopium
(\sect{Microscopium}), I made use of 
former constellations defined in the same area of the sky. This 
was because these former constellations allowed for better 
mythological associations. In Camelopardalis, the giraffe, I used 
Custos Messium, the harvest keeper, defined by Lalande
{\footnote{\url{http://en.wikipedia.org/wiki/Former_constellations}}}, 
that gives a good connection to Ceres. 
 
In Microscopium I preferred to use Globus Aerostaticus (hot air balloon), 
another constellation drawn by Lalande. The balloon reminds us of 
mankind's attempts to fly, and thus to the myth of 
Daedalus and Icarus. 

\section{The planets per constellation}

\subsection{Andromeda}
\label{Andromeda}

Andromeda, named after the Ethiopian princess, offered in 
sacrifice and later saved by Perseus, has an obvious 
mythological association. I suggest the planets around stars 
of that constellation be named after the descendants of Perseus 
and Andromeda. They are seven sons, Cynurus, Perses, Alcaeus, 
Heleus, Mestor, Sthenelus, and Electryon; and two daughter, 
Gorgophone, and Autochtoe.  The dynasty of the Perseus and 
Andromeda, the Perseids, has a profusion of legends where to 
draw further names from. 

\subsection{Antlia}
\label{Antlia}

Antlia, the air pump, is a constellation invented by Lacaille, who, as 
mentioned in \sect{sect:lacaille}, discontinued the tradition of 
mythological names. As explained in that section, I propose to 
name planets in his constellations after inventors in the classical 
myths. I pick here Palamedes for the planet 
of HD 93083. As a fighter in the Trojan War, his myth has a reasonable 
length where to draw more names. Vulcan, Mercury, Minerva, and other 
inventors can be other sources as well. 

\subsection{Apus}
\label{Apus}

Apus, the bird of paradise, is one of the constellations defined by 
Plancius. As a bird, it should be draw association with Diana, the hunter goddess 
and also protector of wildlife in general. The planet around HD 131664 
in Apus could then be called Virbius (Roman counterpart of Hyppolytus), 
a fair hero (or god in the Roman version) who spent his days hunting 
with Diana. 

\subsection{Aquarius}
\label{Aquarius}

The constellation of Aquarius represents Ganymede, 
the cup-bearer of the gods. His family tree therefore provides 
a good source of names for the planets in the constellation. The
planets of Gliese 876 could be named Dardanus, 
Tros, and Ilus, after his great-grandfather, 
father, and brother, the founders of  
Dardania, Tros, and Illium, the three villages that 
amalgamated into Troy. The associated with Troy provides yet 
another source of names for futurely discovered planets. 

Other planets could be called Assaracus (the other brother of 
Ganymede); Themiste, his  niece, daughter of Ilus; Capys, his 
nephew, son of Assaracus and Aigesta (or Themiste, according 
to a variant of the legend). Other related character
is Teucrus, after whom the land Teucria was named. Teucria 
is the former name of Dardania, and is yet another name for the Troad. 

In another variant, the figure in the constellation is Deucalion, the water 
being poured representing the Deluge. I prefer to keep Aquarius as
Ganymede, the more accepted version, and associate Decaulion with 
the constellation of Grus (\sect{Grus})

\subsection{Aquila}
\label{Aquila}

Aquila, the Eagle, appears three times in Greek mythology. First, 
it is the eagle that carried Jupiter's thunderbolts in the ten-year fight 
against his father Saturn and the Titans for the control of 
the world. Later, it
is the eagle (or Jupiter himself in the shape on an eagle) that 
abducted Ganymede to Mount Olympus. In a third appearance, it is 
Ethon, the eagle sent by Jupiter to torture Prometheus by repeatedly
eating his liver day by day. 

As the myth of Ganymede is already used in Aquarius, the latter is 
a more useful source of names, embodied in the fascinating 
myth of Prometheus. Son of Iapetus, one of the Titans, Prometheus 
created mankind from 
clay, also giving them reason. The Olympics did not think much of the new 
integrants of the world, and actually welcomed the rituals of sacrifices and 
honor and reverence to them. The balance, however, tilted when Prometheus 
stole the fire of the gods and gave it to men. Men now did not need to fear 
cold or darkness, they could cook their grains, roast their meat, forge their 
weapons. Suddenly they were less dependent on the gods. The Olympics feared 
that men could eventually take their place on the command of the world, 
as they supplanted the Titans before them. Mankind's paradise had to be
 destroyed, and Prometheus punished for the crime of creating a race 
that rivaled the 
gods. 

Ethon and Prometheus are obvious choices of name. Other is Epimetheus, 
Prometheus brother. Pandora does not strike as a planet's name, but 
Pithos, Pandora's box, could feature, as well as its most precious contents, 
Elpis, hope in ancient Greek. Prometheus was chained 
by Cratos, Bia, and Vulcan, on top of a mount in the Caucasus. 

Bia (force) and Cratos (power) are two of four siblings, the other 
two being Nike (victory) and Zelus (zeal). The former is associated with 
Minerva, and should be reserved for her. Zelus fits, although the 
name closely resembles ``celo'' (jealousy) and ``zelo'' (zeal) in  Spanish 
and  Portuguese, respectively. Of course, there is no surprise, since the 
etymology of both words come from the Greek deity. I also suggest 
to name one of the planets after the mount where 
Prometheus was chained. The mount is usually associated with Mount Elbrus, 
the highest mount in the Caucasus.

\subsection{Ara}
\label{Ara}

Ara, the Altar, is associated with the altar of Lycaon, king of Arcadia, 
who slaughtered and dismembered one of his 50 sons, offering the flesh 
in a banquet to the gods. Jupiter restored the dead son to life, and punished 
Lycaon by striking the remaining 49 with lightning, and changing him into 
a werewolf. There are many different versions of the myth, one states that 
the offered child was Arcas, other that it was Nyctimus. In another version, 
the 49 remaining children were not killed, but also turned into wolves. In 
any case, we have enough names to choose from the kin of Lycaon. For the 
seven planets discovered in Ara, I suggest Pelasgus, Phassus, Nyctimus, 
Peucetis, Caucon, Cynaethus, Stymphalus, Melaeneus, and Eumon.

\subsection{Aries}
\label{Aries}

Aries represents Chrysomallos, the winged ram with golden fleece. It 
features prominently in the myth of Jason and the Argonauts, as the 
main goal of the expedition. As I reserve most of the myth for the 
constellation of Puppis, Vela, and Carina (the parts of the Argo), 
here I use only part of the myth of Aries that not associated with 
the Argonauts. As with Pegasus and Bellerophon, it is natural to associated 
a planet with the character(s) who rode Aries. In this case, it is Phrixus, 
son of Athamas and Nephele, the cloud goddess made by Jupiter in the shape 
of Juno, also mother of the centaurs. Nephele sent Aries to 
save Phrixus and his twin sister, Helle, from their evil stepmother Ino, 
who plotted to kill them. Phrixus flew to Colchis, where he was adopted by 
king Aeetes. There he married Chalciope, daughter of Aeetes, and sacrificed 
Aries to Jupiter.

We already have in this paragraph six names for the planets found in Aries 
to date. The planets in the triple system around HIP 14810 can be named
Phrixus, Helle, and Chrysomallos, thus the planets revolving around the star 
share a mythological resemblance with the flight of the twins on the back of the 
winged ram. In the myth, Helle fell and drowned in the Dardanelles, which is 
where the ancient name of the strait, Hellespont, comes from. HIP 14810d, the 
lighter and most eccentric of the three planets, can be named after her. 

The planets 
around HD 12661 can be named Aeetes and Chalciope, 
after Phrixus new family in Colchis, while HD 20367 b can be named Colchis 
itself. The names Ino and Athamas can be reserved for further planets found 
in the constellation. Nephele better fits with the centaurs, her more famous 
offspring. 

\subsection{Auriga}
\label{Auriga}

Auriga, the charioteer, usually represents the blacksmith god Vulcan.
Planet Vulcan has become sort of a running gag in astronomy. It was the name given 
to a hypothetical planet supposed to exist in an intra-Mercurian orbit, the inner 
solar system equivalent to Planet X (not Pluto). Before Einstein's general
relativity, a hypothetical planet was the favored explanation 
for the precession of Mercury's perihelion. Le Verrier called it Vulcan and 
calculated its orbit. His success with Neptune prompted a search and  
soon transits of Vulcan were reported. Le Verrier died convinced that he 
had discovered yet another planet, but in the end, no conclusively 
evidence of Vulcan could be found. The name nevertheless stuck with both 
with the general public and among astronomers, since the blacksmith god is 
such a good name for an object so close to a star. The former included 
a planet Vulcan in the series Star Trek, while the latter are still looking 
for hypothetical Vulcanoid asteroids between the Sun and Mercury. It was 
even suggested that the whole class of Hot Jupiters should be called 
``Vulcan Planets''. It seems that we are just dying to use the name.

So here it is. The transiting planet WASP-12 b seems the best proxy 
of the class of Hot Jupiters among the planets discovered in Auriga. 
With a mass of 1.41 $M_J$, and a semi-major axis 0.0229 AU, it circles the star in 
1.09 day. This scorched hot planet in Auriga certainly deserves the title of 
Vulcan. Since his beautiful wife Venus already lies in the 
Solar system, it seems fair that the smith god should have the benefit of Aglaia, 
the youngest of the three Graces, who is attributed to be his wife in a variant 
of the myth. Lemnos, the place where he landed when thrown from 
Olympus by his sweet of a mother; and Lycia, the place where his cult originated, 
are also good choices. Euthenia, one of his daughters with Aglaia closes 
the list of names for Auriga.

\subsection{Bootes} 
\label{Bootes}

The constellation of Bootes, or herdsman, has no clear representation. 
In Roman-Greek mythology he is Arcas, the son of Callisto and Jupiter, 
whereas others interpret it as being Icarius (not Icarus), a herdsman who 
was taught the art of wine-making by Bacchus himself. Yet another 
interpretation has him as Atlas, who carried the 
world on his shoulders. The ambiguity is welcome, since the constellation 
is large. Arcas is also said to be Ursa Minor since, in a variant of the 
myth, Jupiter transformed both Callisto and Arcas into bears. Arcas lends
its name to Arcadia, region of Greece famous for its bucolic lifestyle. 
Arcadia is also where the cult of Apollo flourished, as well as the 
region where Mount Cyllene, the birthplace of Mercury, stands. 
Mercury's mortal mother, Maia, also raised Arcas 
in a variant of the myth. Since Bacchus already appears in the constellation, 
we save Mercury for Lyra, the lyre, the instrument that he invented. Apollo 
has no clear strong association with any constellation (and loose 
associations with many), so I refrain from using his myth here as well. 

I propose to name the planets of Bootis after 
Arcas (tau Boo b), Atlas (HD 128311 b), and Bacchus (HD 132406 b). 
Other names are Arcadia for WASP-14 b, and Pramnos for HAT-P-4 b. 
The second planet of HD 128311 could be called Aithra, after 
the Oceanid nymph with whom Atlas begot the Hyades. 

Atlas is purposely included. It is already a name given to 
a star of the Pleiades, a satellite of Saturn, and a crater 
on the Moon. It illustrates that the same designation does not 
cause confusion since the objects pertain to different fields of 
study.

\subsection{Caelum}
\label{Caelum}

Caelum, the chisel, contains no known planet-hosting stars as of Oct 2009. 
As an art-related constellations, names can be drawn from Apollo's myth.

\subsection{Camelopardalis}  
\label{Camelopardalis}

Camelopardalis, the giraffe, is a constellation created by Plancius. As a 
land animal, names for planets should be drawn from Diana's myth. 

Egeria, a nymph associated with Diana. Triklaria, one of her titles, and 
Ephesia, after her main local of adoration, may as well appear as 
names for planets in this constellation.

A loose association can be established with Ceres or other 
agricultural deities, as mentioned in \sect{sect:former}. I therefore 
suggest one of the planets be called Opalia, after the festivities 
to the goddess Ops. Ops, mother of Ceres, is the Roman equivalent to 
Rhea, mother of Demeter (Cybele was actually her name in Phrigya). 

\subsection{Cancer}
\label{Cancer}    

Cancer, the crab, plays a minor role in Greek-Roman mythology, 
namely, in Hercules' twelve labors. While Hercules was fighting 
the Hydra of Lerna, Juno sent a crab 
to distract him. Hercules simply crushed the insignificant 
creature. Grateful for the crab's effort, Juno gave it a place in sky.

The connection with the Hydra allows for 
associating Cancer with the region of Lerna, full 
of mythological detail. The Danaids, for instance, 
buried in Lerna the heads of their husbands. I take 
three names from the Danaids, Anthelea, 
Stygne, and Euippe. From Juno 
we may take two of her titles, Argive, and Teleia, 

For the planet recently discovered around HD 73534, I draw again 
from the Danaids, and suggest the name of Pirene.


\subsection{Canes Venatici}  
\label{Canes Venatici}  

Canes Venatici, the hunting hounds, is a spurious constellation. 
Historically part of Bootis, it was mistranslated from Greek 
(as cudgel) to Arabic (hook) and once again from Arabic back to 
Latin (dogs). Hevelius (1690) formalized them as Bootes' 
hunting hounds, Asterion and Chara. Asterion is now known as 
Cor Caroli, which releases the name to planetary use. The only 
planet-hosting star in the constellation is HAT-P-12. I propose its 
planet be called Asterion, after the old name of Cor Caroli.

\subsection{Canis Major}     
\label{Canis Major}     

Canis Major represents the dogs of Orion, the mighty hunter. One source 
cites them as Leucomelaena, Maera, Dromis, Cisseta, Lampuris, Lycoctonus, 
Ptoophagus, and Arctophonus. We get the first six names. The constellation 
has a profusion of variants, maybe representing Laelaps, the mythological 
dog who never failed to catch a prey; maybe the hound of Procris, a nymph 
of Diana; or the mightily fast dog given by Aurora to Cephalus. Laelaps 
however, appears more prominently in the myth of the Teumessian fox, and 
therefore we choose to use it for a planet in Vulpecula (\sect{Vulpecula}). In any 
case, there are plenty of sources for further names in Canis Major, that 
can as well be applied to Canis Minor, when planets are found there. 

\subsection{Canis Minor}     
\label{Canis Minor}     
Canis Minor contains no known planet-hosting stars as of Oct 2009.

\subsection{Capricornus}     
\label{Capricornus}     

The constellation of Capricornus, the goat, may represent Pan, 
the powerful Faun, or Amalthea, the goat that nourished the 
infant Jupiter. I prefer to 
connect the latter with Monoceros, that lacks other major mythological 
associations, and use Pan for Capricornus. 

Pan is a pastoral deity, guardian of flocks and shepherds. His father 
was Mercury, who in the shape of a goat conceived him with a 
doubtful mother. The nymphs Dryope and Oeneis; Penelope, the wife of 
Ulysses; and even Amalthea herself, are mentioned in different versions 
of the myth. As a Faun, Pan is depicted as a man with horns, tail 
and feet of a goat. He lived among the nymphs, and claimed to have 
seduced many of them. One, Syrinx, was not so interested and fled 
in terror. She was turned into a clump of reeds, from which Pan 
made a pipe, syrinx, the pan flute. Echo, usually associated with Narcissus, 
was also loved by Pan in a variant of her myth. 

The connection between Pan and Capricornus comes from the episode where he leapt 
into the Nile to escape Typhon as Jupiter struggled with the monster. 
His head became that of a goat, and his hindquarters the rear part of a 
fish. He was later elevated to the skies as Capricornus. The constellation 
is referred sometimes to as the ``sea-goat'' because of it. But the name 
in Latin, capri-cornus, translates simply as goat-horn. 

I propose to name the three planets known in Capricornus Syrinx, Echo, and 
Dryope. Pan itself is not used since it is already a moon of Saturn. 
Actually, the main reason is subjective. Pan is homophonous to the 
widely used Greek word {\it pan}, to which the god 
has no connection. Echo has the same name as the acoustic phenomenon, but 
here there is an immediate connection since the phenomenon was named after 
the nymph or vice-versa. 

\subsection{Carina}
\label{Carina}

Carina, like Puppis and Vela, is a part of the ancient constellation of 
Argo Navis, representing the ship of Jason and the Argonauts. Due to 
its immense size, Navis was divided by Lacaille (1763) into Carina, the 
keel, Puppis, the poop deck, and Vela, the sails. I will refer to the 
three of them as Navis or Ship. 
The myth can be no other than the myth of the Argonauts, the 50-60 heroes 
who boarded the Argo with Jason in his quest for Aries, the golden Fleece. 

I include the name Jason, even though it is a common 
male name in English-speaking countries (I add though, that the 
original pronouncing is {\it Yasson}). Other planets in Carina are 
obviously named after the Argonauts. Puppis and Vela will 
naturally draw from the same source. 

As the Ship takes a very large fraction of the sky, we expect 
many planets to be found there. Other sailing myths such as the 
Iliad and the Lusiad may be used in the future.

\subsection{Cassiopeia}      
\label{Cassiopeia}      

Cassiopeia, queen of Ethiopia, once unwisely bragged that 
she and her daughter Andromeda were fairer than the Nereids. 
This angered Neptune, that put her near the pole, where 
she would spent half of her time upside down. I propose 
to name the planets in Cassiopeia after the Nereids, to 
better torment the poor vain queen. Eulimene, 
Orinthya, and Thetis are my choices for the three known 
planets.

\subsection{Centaurus}
\label{Centaurus}

The centaurs are well known to astronomers, as the minor bodies 
between the main belt and the Kuiper belt. As most of them 
are already designated after a mythological centaur, we have no choice 
but to used repeated names. However, I reserve Chiron for Sagittarius, the 
centaur usually associated with that constellation. In his place 
I include Nephele, the mother of the Centaurs. 
       
\subsection{Cepheus}         
\label{Cepheus}         

Cepheus, Andromeda's father, is a constellation where just one 
planet has been confirmed. I propose to name it after Dannaus, 
Cepheus' brother. 

\subsection{Cetus}
\label{Cetus}

Cetus, the whale, is the sea monster sent by Neptune to 
terrorize the coast of Ethiopia in order to punish Cassiopeia for her 
arrogance. Cetus plays a non-negligible role in Greek 
mythology as she and her consort Phorcyd, a primordial 
sea god, sprang many other monsters, collectively called 
the Phorcyds. I propose they should name the planets 
circling stars in that constellation. Scylla, a six-headed 
monster; Stheno and Euryale, two of the Gorgons (the other 
being Medusa); Echidna, usually the mother of all monsters, but 
according to a variant of the legend also an offspring of Cetus;
Deino, one of the Graeae, three horrible sisters that shared 
one eye and one tooth among them; and Thoosa, mother of the cyclops
Polyphemus.  
Scylla also naturally brings Charybdis, the sea monster with whom 
she teams up on the task of dooming unadverted sailors. 

Cetus is a part of the sky called the Sea, for the profusion of 
water-related constellations. As such, I 
strongly encourage the name of Ulysses for one of its planets, in 
honor of the hero who navigated through all these 
dangers. However, I use his Greek name Odysseus, since Ulysses 
is a semi-common male name. Planet Odysseus could be a planet of 
one of the double systems, HD 11964, so that the 
other planet is named after his son, Telemachus. The other double system, 
HD 11506, naturally goes to Scylla and Charybdis. The two most recent 
ones can be named after Callidice, wife of Odysseus during his 
voyage to Thesprotia, and Polypoites, their son. The presence of 
Odysseus also allows for the addition of characters of the 
Odyssey as more planets are discovered in the Sea. This area of the 
sky includes Cetus, Aquarius, Pisces, Piscis Austrinus, Eridanus, 
Delphinus, and Hydra. Some also include Navis, Crater, and Capricornus. 

\subsection{Chamaeleon}      
\label{Chamaeleon}      

Chamaeleon is one of Plancius' constellations. Being an 
animal, it welcomes a connection with Diana. However, I am inclined 
to first associate it with Proteus, son of Neptune. Proteus can tell
the future, but will only tell it to someone who is capable of capturing him.
To avoid that, he changes his shape. 

Although shape-shifting 
is a common theme in Greek mythology, Proteus is the one deity mostly 
associated with it, as {\it protean} came to mean ``versatile'', and 
carries a positive connection of flexibility, versatility and adaptability, 
much in the same way as 
someone can be described as a ``chameleon''. It also allows to use the 
myth of Neptune, quite underrepresented so far, instead of Diana, 
who will be shared among many animal-related constellations. 
Nereus also has the ability to shape-shift, and 
may be used as well. He also figures in a version of the myth as Proteus 
father, the mother being a Naiad, nymphs of springs and fresh water. 

As said in \sect{sect:naming}, I refrain from using the names of major moons. But 
Naiad is only a smaller moon of Neptune. It may as well 
figure as a planet with no major source of confusion. 

\subsection{Circinus}
\label{Circinus}

Circinus, the drafting compass, contains no known 
planet-hosting stars as of Oct 2009.

\subsection{Columba}        
\label{Columba}        

Columba, the dove, is one of Plancius' constellations. 
He named it after the dove of Noah, 
that gave him the information that the Flood had stopped.  A flood 
legend also figures in Greek mythology, when Jupiter decided to 
end the Bronze Age and sent the Deluge. However, I associate that 
story with Grus (\sect{Grus}), and prefer to associate Columba with
a passage of the Lusiad. Cam\~oes describes the chariot of 
Venus as being pulled by swans and surrounded by doves who
playfully circle it. He names at least one, Peristera{\footnote{As English is 
language with a penchant for paroxitones, I explicitly inform 
that Peristera is a proparoxitone name, which, of course, makes 
it more beautiful.}}, a nymph converted 
into a dove by Cupid. That fits well for the only planet discovered 
so far circling a star in Columba (though the ``planet'' actually 
seems to be a brown dwarf). Further discoveries may 
drawn from the myths of Venus and/or Cupid, like Pisces (\sect{Pisces}). 

\subsection{Coma Berenices} 
\label{Coma Berenices} 

Coma Berenices, or Berenice's Hair, is one of the few constellations that 
is named after a historical rather than mythological figure, Queen
 Berenice II of Egypt. However, Eratosthenes referred to it as both 
Berenice's Hair or Ariadne's Hair. I take the latter to keep the 
associations mythological. 

Ariadne is the daughter of Minos, king of Crete, who ordered the 
construction of the Labyrinth to hold the Minotaur. This immediately 
associates the constellation with the myth of Daedalus and Icarus, 
the Labyrinth's most famous occupants after only the Minotaur himself, and 
perhaps Theseus, who killed the Minotaur. In the more common version of the 
myth, Ariadne fell in love with Theseus. In a variant, she is the bride 
of Bacchus. 

Two stars in Coma Berenices are known to have substellar companions,
though one of them, HD 114762, is probably circled by a brown dwarf. 
The other star, HD 108874, harbors a double system. The double system 
may be called Ariadne and Theseus. The massive planet can be called 
Naxos, after the island where Bacchus met Ariadne. HD 108874 b is 
a planet with close Earth-like insolation, so the name of the fair 
Ariadne is quite fit. 

\subsection{Corona Austrina}
\label{Corona Austrina}
Corona Austrina, the southern crown, contains no known planet-hosting stars as of Oct 2009.

\subsection{Corona Borealis}
\label{Corona Borealis}

Corona Borealis is associated with a crown that Bacchus gave to Ariadne. 
Three stars with planets are known there, that I suggest be named 
after three of their children, Euanthes, Staphylus, and Latramys.

\subsection{Corvus}
\label{Corvus}
Corvus, the crow, is the bird of Apollo. In the myth the crow as a speaking 
bird with white feathers, and loyal to the god. The bird was put in charge 
of watching over Apollo's love, Coronis, who was then pregnant with Asclepius 
(see \sect{Ophiuchus}). The crow witnessed Coronis being unfaithful to him 
with a mortal, Ischys, and reported it to Apollo. The god was furious, and 
unjustly turned his anger on the unfortunate bird, scorching his feathers 
black and removing its ability to speak. He later also had his sister Diana 
kill Coronis because he could not bear doing it himself. 

As of Oct 2009, there is one known planet circling a star in Corvus. I suggest 
naming it Coronis. 

\subsection{Crater}         
\label{Crater}         

Crater, the cup, represents the cup of Apollo. The legend does not extend 
beyond a couple of lines. Corvus serves him water, but lazily brings it with a water 
snake inside. Apollo angrily throws them all into the sky.

With little to draw from, we may as well associate 
the constellation with Ganymede again, or Hebe, the 
cupbearer before him. With Hercules, Hebe had two sons, Alexiares and Aniketus,
gatekeepers of Olympus, who may lend their names to the two planets 
discovered so far in Crater. Hebe herself is too common a name.  

\subsection{Crux}
\label{Crux}

Crux, the Southern Cross, is a constellation of major significance for 
navigation in the southern hemisphere. Unlike in the northern hemisphere, the 
celestial south pole has no bright star to mark its position. However, 
we can rely on the Southern Cross to point our direction. Its major arm, 
prolonged 4.5 times, marks the position of the celestial south pole. It 
is also of cultural significance, appearing in the flags of five countries, 
namely, Australia, Brazil, New Zealand, Papua New Guinea, and Samoa. In Brazil,
 where it is known as {\it Cruzeiro do Sul} or simply {\it Cruzeiro}, 
it even featured as the name of the currency (from 1942 to 1986 and again 
from 1990 to 1994), as well as of a football team. These facts underscore 
that its main association is not with 
Christianity, but with {\it south}. 

Two planets are known in Crux, around 
the stars HD 108147 b, and NGC 4349 No 127 (that a star in a cluster). 
I suggest to name them after Livas, the Southwest wind, and Apeliotes, 
the southeast wind. The south wind Noto is used for a planet in Octans, the 
constellation that contains the south pole.

\subsection{Cygnus}         
\label{Cygnus}         

Cygnus, the swan, was the disguise used by Zeus to seduce and impregnate 
Leda, a much used theme in Renaissance art. I suggest the planets of 
Cygnus be named after Leda and her family. Thestius, her father; Iphicles, 
Eurypylus, two of her brothers; Althaea, her sister; Timandra, her daughter, 
and Echemus, Timandra's husband may name the double system of HD 187123. 
A futurely discovered planet may be called after Cycnus, name of three 
characters of Greek mythology who were transformed into swans.

\subsection{Delphinus}      
\label{Delphinus}      

Delphinus, the dolphin, has a minor role in Greek mythology, associated 
with Arion, a poet of Lesbos allegedly from the 7th century BC. Although 
his historical existence is a matter of ongoing debate, legendary 
for sure is the story of his kidnapping, that rendered the association 
with the dolphin. Sailing back home after winning a music competition, 
the crew of the ship plotted to kill him and steal the prize. His last 
wish was to play one last song, a hymn to Apollo. The beautiful song 
attracted dolphins, and he jumped to the sea, being saved by one of them. 
The story is perhaps inspired in the myth of Melikertes, also associated 
with the lesser known Roman deity Portunes, god of harbors and ports.  

The myth allows for a loose association with Apollo. Although some might 
argue that Apollo as a solar deity should figure in a zodiacal constellation, 
the resemblance of the names Delphinus and Delphi, the Oracle of Apollo 
and the most famous in Greek mythology, somehow suggests the association. 
I therefore suggest the four planets known in Delphinus to be called 
Delphi, Apollo, Melikertes, and Portunes.  

\subsection{Dorado}         
\label{Dorado}         


Dorado is one of the constellations created by Plancius. Dorado or Dourado 
is the name given to many species of fishes, the best known of them being 
the goldfish (``dorado'' literally means ``golden'' in Spanish). It is also 
a major fish of the Amazon. There are two known planets circling stars in 
Dorado. Being associated with water, I suggest one of them be named Tyro,  
lover of Enipeus, a river-god. She was also one of the many love 
adventures of Neptune. The other may as well be named Enipeus. 

\subsection{Draco}          
\label{Draco}          

Draco is a constellation representing Ladon, the hundred-headed dragon who 
guarded the garden of the Hesperides, nymphs of the evening, 
prominently featured as the 11th labor of Hercules. The names and 
number of the Hesperides varies according to variants of the myth. 
I list Aegle, Erysteis, Lipara, and Chrysothemis as possible names. Another 
of the five planets discovered circling a star in Draco can of course be named 
Ladon. 

\subsection{Equuleus}
\label{Equuleus}

Equuleus, the little horse, contains no known planet-hosting stars as of Oct 2009.
\subsection{Eridanus}
\label{Eridanus}

Eridanus in most versions of the Greek myths is a river that 
surrounds the world. Virgil, however, lists it as one of the 
rivers of Hades, the underworld, realm of Pluto. Hades has other 
five rivers, Acheron, Cocytus, Phlegethon, Lethe, and Styx. These 
are obvious choices for planets around the stars of Eridanus. 
Further names can be drawn from the myth of Pluto. In particular, 
instead of Lethe I include Radhamantus, one of the three judges 
of Hades. Eachus, other judge, is also included (the third judge is 
Minos). 

\subsection{Fornax}   
\label{Fornax}   

Fornax, the furnace, is one of Lacaille's constellations. Being 
a furnace, it is immediately associated with Vulcan. Aetna, or Etna, 
after the mountain where his workshop was supposedly located, invites 
for a naming after volcanos. Lipari, for the association 
with Vulcan. Milos, a volcano of Greece, and homonomous volcanic 
island, is welcome since it is also associated with the famous statues 
Venus of Milo, thus indirectly connecting Vulcan and Venus, as in the 
myth. 

\subsection{Gemini}  
\label{Gemini}  

Gemini, represent the twins Castor and Pollux, the dioscuri, sons of Leda. 
As both are already names of stars, I suggest the names of the other famous 
twins, Romulus and Remus, the mythical founders of Rome. The third planet 
in Gemini can be Lupa, the she-wolf that nourished the twins. 
       
\subsection{Grus}
\label{Grus}

Grus, the bird crane, is one of Plancius' constellations.
The crane appears once in classic mythology, associated 
with the Greek version of the flood, as already mentioned in Columba. 
When Jupiter was about the send the deluge, Prometheus advised his son 
Deucalion to build an ark. He and his wife Pyrrha were thus saved from the 
rising waters. When the deluge was over, Deucalion and Pyrrha consulted the 
oracle of Themis on how to repopulate the Earth. They were told to 
throw the bones of their mother over their shoulders. Pyrrha 
interpret the mother as being the Earth, the bones being stones. 
The stones Deucalion threw became men; the ones that Pyrrha threw become 
women. 

Other men escaped from perishing in the deluge as well, by clinging 
to the top of high mountains. One of them was Megarus, who swam to the top 
of Mount Gerania, following the sound of cranes. There are four planets 
discovered around stars in Grus, that I suggest should be named
Deucalion, Pyrrha, Megarus, and Gerania. 

\subsection{Hercules}       
\label{Hercules}       

Hercules, as one of the richest myths of ancient Greece, should be 
a plentiful source of names on its own. 

In honor of the herculeous 
effort of those who discovered the exoplanets, I propose to name 
these planets after associations with the twelve labors of the hero.
The first two, the Nemean Lion and the Lernaean Hydra, already have 
associated constellations, and should pertain there. But the other 
ten are there for the taking. I therefore suggest Cerenytis, 
Erymanthus, Augean, Alpheus, Peneus, Stymphalia, Diomedes, Geryon, and Cerberus.

It is worth noting that although Cerberus could figure in Eridanus, 
associated with Hades, it also fits well in Hercules. The reason is 
that Cerberus was a former constellation, defined by Hevelius (1690), who
envisioned Hercules in the sky struggling with the three-headed dog. 

\subsection{Horologium}     
\label{Horologium}     

Horologium, the hourglass, is a modern constellation. As it associated 
with time, I can think of no other but Cronus, Saturn's Greek name, for 
the planet of HR 810, Iota Horologii, the only known planet-hosting star
in that constellation. 

\subsection{Hydra}          
\label{Hydra}          

The constellation of Hydra, representing the Lernaean Hydra, deserves some 
pause. Hydra figures only as the 2nd labor of Hercules, yet it is the 
biggest of the 88 constellations. Even though Lerna is 
full of mythological detail, I feel forced to 
also use the constellation's association with the Sea, in 
order to maximize the possibilities 
of names for the constellation. So it is that apart from 
Lerna, I summon first some Nereids. Amphitrite, Galatea, Pasithea, 
Nausithoe, Menippe, Thaleia, Spio, Ianira, and Asia. Yes, Asia. The continent is 
named after her, just like Europe is named after Europa, another nymph, 
that also named the second Galilean moon. Also considering the 
difference in size between the 
continents, it sounds reasonable that if a moon is named after Europa, 
that a planet should be named Asia. Lest someone gets excited on the other 
side of the Atlantic, Vespucci was a sea man, not a sea beauty. 

As said in Cancer, Lerna has other mythological significances. The 
Danaids, for instance, 
buried in Lerna the heads of their husbands. To complete the naming of the 
known planets in Hydra, we may use three Danaids, Adiante, Amymone, and 
Hyperippe. Amymone
also resonates with the Hydra itself, since the monster had its lair in the 
spring of Amymone, deep in a cave in Lerna. 

Another trick, need it be, and already used in Cancer, is to associate 
it with Juno. After all, it was her who forced Hercules to execute 
the labors. 

\subsection{Hydrus}  
\label{Hydrus}  

Hydrus, the water snake, is a new constellation, but associated with the 
water snake that Corvus brought to Apollo in his cup (see Crater). Apollo 
angrily threw all of them to the sky. Being nothing but a small annoyance
to the god, 
we may as well name planets in these constellations after another 
particular nuisance. Delos and Ortygia, his and Diana's places 
of birth may represent 
such difficulties. Juno, irritated with yet another love adventure of Jupiter,
kidnapped Ilithyia, goddess of childbirth, in order to prevent Leto from 
going into labor. 
       
\subsection{Indus}          
\label{Indus}          

Indus, the Indian, is one of Plancius' constellations. It represents
an Indian, by the time referred to either a native of India or of the 
Americas. I suggest Hesper, the evening, the setting sun, and thus
the Western Hemisphere. The same name is used by Joel Barlow to represent 
the lands of the American continent in the Columbiad.

\subsection{Lacerta}      
\label{Lacerta}      

Lacerta, the lizard, has one planet discovered. I suggest it be named 
after Abas, son of king Celeus and prince of Eleusis (see Linx, \sect{Lynx}), 
who was transformed into a lizard by Ceres. Further names can be drawn from Ceres' myth. 

\subsection{Leo} 
\label{Leo} 

Leo represents the Nemean lion, Hercules first labor. It may also 
be related to Bacchus, since the lion was an animal closely associated 
with the wine-god. Omphale, who wore the skin of the lion, Lamus, her son 
with Hercules; and Tmolus, of Omphale. Naturally, 
Nemea;  Elissos, a river in Nemea; Iraklion, 
former name of Nemea; Lycurgus, kind of Nemea; and Cleoane, 
near where the Nemean Games took place, 

Nemea also figures in the myth of the Seven Against Thebes (Aeschylus 467BC), 
concerning the battle between an Argive army led by Polynices and the army 
of Thebes. It is related to the myth of Oedipus and Jocasta, and a good 
source of future names for Leo and Leo Minor. It was written as a play, 
and reportedly won the first prize at the City Dyonisia, a large religious 
festival in ancient Athens in honor of Bacchus. The association 
with Leo is simple yet sufficient: the Seven pass by Nemea on their way 
to Thebes. The occurence was not exactly uneventful. It even resulted in a 
fatality, as described below in Leo Minor. 

\subsection{Leo Minor}
\label{Leo Minor}

Leo minor, a lion cub accompanying Leo, has no myth of its own, 
being completely 
correlated with Leo. I suggest the name Archemoros for the planet 
recently discovered around HD 87883. Archemoros was a infant prince of 
Nemea who died strangled by a snake while his nanny Hypsipyle 
was off to fetch water to the Seven. 

\subsection{Lepus}        
\label{Lepus}        

Lepus, the hare, is the favorite prey of Orion and is constantly being 
hunted by him in the sky. I suggest that the planet around HD 33283 
could go by the name of Epimelius. It is one of the many titles of 
Mercury, Hermes Epimelius, meaning keeper of flocks. 
Even though the hare is a not a flock animal, 
Epimelius can be thought to highlight Mercury's animal welfare attributes
in general. Being so routinely hunted by Orion, poor Lepus may be in need 
of some divine protection.

\subsection{Libra}
\label{Libra}

Libra, the weighting scale, can most obviously be associated with Themis, 
the Titanian who personificates Justice. I therefore propose to 
name the planets around Gliese 581 after her and the Horae, her daughters 
with Jupiter. These are Eirene, Dike, and Eunomia, as mentioned in an 
example in the introduction. Themis and Jupiter also fathered Astraea, 
the star-maiden, yet another personification of Justice. In a variant of 
the interpretation of the constellations, she is Virgo, and Libra the 
scales she carries. I do like to include her in Libra, but Virgo may 
be associated with a multitude of other goddesses of more relevance. The 
presence of the Horae also allows for an association with a different 
set of Horae, the Hours, allegories for the hours of the day. Hesper, 
already mentioned in Indus (\sect{Indus}), is one of them, 
representing Evening. We may as well pick Dysis, the Sunset. 

\subsection{Lupus}        
\label{Lupus}        

Lupus, the wolf, may represent Lycaon, the werewolf, already 
mentioned in Ara. It was named Therion by Hipparchus (meaning beast). Its 
two known planets may therefore be named Lycaon and Therion, and further 
names be used from the myth of Lycaon or other generic beasts of mythology.  
From Norse mythology, Fenris can be used, after the wolf that 
eats Odin in the Ragnarok.

\subsection{Lynx}
\label{Lynx}

Lynx is one of Hevelius' constellations. Although not named 
after a mythological figure, we can nevertheless draw an association 
with Lyncus, the king of Scythia, who was transformed into a lynx by 
Ceres. 

When Demeter was looking for Persephone, having taken 
the form of an old woman called Doso, she received a hospitable welcome 
from Celeus, king of Eleusis. He asked her to nurse Demophon and Triptolemus, 
his sons by Metanira. As a gift to Celeus, because of his hospitality, 
Demeter planned to make Demophon immortal by burning away his mortal spirit 
in the family hearth every night. She was unable to complete the ritual 
because Metanira walked in on her one night. Instead, Demeter chose to 
teach Triptolemus the art of agriculture and, from him, the rest of 
Greece learned to plant and reap crops. 

Lyncus, however, did not want to learn the arts, and tried instead to 
kill Triptolemus. As a punishment, Ceres turned Lyncus into a lynx. 

\subsection{Lyra}
\label{Lyra}

Lyra, the lyre, appears in many legends, since it is the main 
musical instrument of the Antiquity. According to the myth, 
the lyre was invented by Mercury, not long after his birth. 
The young god stole the cattle of Apollo, and sacrificed one. 
Making strings out of the stretched entrails of the animal and 
the arms from the horns, he fashioned the first lyre. Apollo went
furious when he noticed the robbery, and went to Mount Cyllene 
with the intent to punish the trickster. However, Mercury 
played the lyre, and Apollo got marveled by the sweet sound of 
the instrument. He forgave Mercury in trade for the lyre, that 
became one of his  main attributes. Indeed, Apollo's importance 
as patron of music, poetry, and arts predates his later 
association with Helios as sun-god. 

As for mortals, the lyre figures most prominently in the myth of Orpheus. 
By some considered son of Apollo, Orpheus was taught the instrument by the 
Muses, and played it to perfection. Jason had him on board of the Argo on 
advice of Chiron, who said that he would be needed if the Argonauts were 
ever to pass the Sirens. The Sirens were sea monsters that, disguised 
as nymphs, played so sweet music that the seafarers were enchanted and lured 
to their deaths. When the Argo did pass by them, Orpheus drew his lyre. 
His music was more beautiful than that of the Sirens; so sublime that the 
bewitching chords went unnoticed by the Argonauts. Orpheus' most known myth, 
however, is his descent into Hades in the failed attempt to bring 
back his beloved Eurydice.  

The association with Apollo is tempting. However, we are running out of 
names, and Lyra already has Mercury, Orpheus, and the Sirens as sound 
sources. Furthermore, Apollo has loose associations with many other 
constellations, some of which have him as only well-developed myth  
where to draw names from, as in the case of Delphinus. I therefore 
propose to name the planets in Lyra after Hermes, Mercury's Greek name; 
naturally Orpheus and Eurydice; and Siren. As the lyre is an instrument, 
other names could be Alipes, the winged sandals of Mercury, and Petasus, 
Mercury's winged hat.

\subsection{Mensa}        
\label{Mensa}        

Mensa, the table, is one of Lacaille's constellations, named 
after Table Mountain near Cape Town, in South Africa, 
where he spent time observing the southern sky. 

This brings an immediate association with Adamastor, the Gigante 
brother of Enceladus who guards the Cape of Storms in the 
Lusiad. Adamastor appears 
as a stormy cloud, sinks the ships that try to round the cape, and dissipates 
into tears, which are the salty waters of the confluence of the Atlantic 
and the Indian Oceans. Also known as the Spirit of the Cape, 
his name is a near-anagram of Cape of Storms in Portuguese, 
{\it Cabo das Tormentas}, with suppression of letters. 
The constellation of Mensa and its association with 
Cape Town seems a good place for Adamastor, who represents the 
dangers sailors faced when trying to round the cape. May he now 
also represent that the stormy challenges of discovery of exoplanets 
were also turned into good hope. 

\subsection{Microscopium} 
\label{Microscopium} 

Microscopium is one of Lacaille's constellations, with no mythological 
association. However, Lalande later tried to rename it to 
Globus Aerostaticus (hot air balloon). As such, it immediately refers to Daedalus, 
who attempted flight in order to 
escape the labyrinth he himself erected. Daedalus is also an inventor, 
so the association with one of the constellations of Lacaille is 
natural. 

However, we may reserve Daedalus for a double system, so that the other 
planet could be called Icarus. The one in Coma Berenice seemed a good choice, 
since Ariadne is associated with the myth of the Minotaur. Instead, 
the double system of CoRot-7 seems better for physical reasons, since 
CoRot-7 b is the planet with smallest 
semi-major axis known to date. The planet is also 
rocky (Queloz et al. 2009), but the 
temperature on the substellar point is so high (1800-2600\,K, 
L\'eger et al. 2009) that its surface is probably molten on the day side. 
The name of the boy who flew too close to the Sun 
while wearing meltable wax wings is a suitable name for such a 
planet. Due to this, I prefer to abandon Coma Berenice and Microscopium 
(or the Balloon, for that matter), in favor of loosely associating 
Daedalus and Icarus with Monoceros. 

Planets around Microscopium-Balloon may therefore be named after winged 
characters of Greek mythology. Alternatively, Perdix (or Talos), Daedalus 
nephew, may as well figure. He was pushed by Daedalus from a tower, 
but Minerva saved him by transforming him into the bird of same name. 
This notorious bird does not build nests on trees. Mindful of falls, 
it avoids high places. It may represent successful, yet careful, flight, 
as opposed to the failed and fatal flights of Daedalus and Icarus.

\subsection{Monoceros} 
\label{Monoceros}

Monoceros, the unicorn, is one of Plancius' constellations. Unicorns 
do not appear in Roman-Greek mythology, but in treaties of natural 
history of the ancient Greeks. Ctesias of Cnidos (5th century BC) 
appears to be the first to mention them. 

A loose association with mythology may be drawn from the myth of 
the Cornucopia, the horn of plenty. Cornucopia is the broken 
horn of Amalthea, the goat that nourished the infant Jupiter, as mentioned 
in Capricornus (\sect{Capricornus}). As a single-horned goat, Amalthea can be said to 
be the Greek version of the Unicorn. Although in this case, the magic 
horn is the missing one.  

In Metamorphosis, Ovid has Achelous narrate to Theseus his fight 
with Hercules for the possession of Dejanira. He transformed 
himself into a bull to fight the 
semi-god, but Hercules tore off one of his horns. He trade it with 
the hero for the horn of Amalthea, which 
Hercules later gave to the Naiads, the nymphs of fountains and springs. They, 
in turn, transformed the horn into Cornucopia. It was a 
magic horn overflowing with fruits and 
grains. Because of this, it is associated with several agricultural 
deities, such as Gaia, Cybele, Ceres, and Pluto; as well as with the 
several river gods and nymphs, as fertilizers of the land. 

The fact that Achelous narrates the story to Theseus reminds us 
of the Minotaur, and hence a loose association with the labyrinth, Daedalus,  
and Icarus. These can be the names of the planets around CoRoT-7, as 
explained in \sect{Microscopium}. 

For the other planets in Monoceros, I therefore suggest the names Cornucopia; 
Achelous; Adamanthea, other name of Amalthea; and Cybele, Jupiter's mother 
Rhea by another name; and Dejanira, who prompted the fight between Hercules 
and Achelous. Also included is Dexamenus, Dejanira's father in a version 
of her myth. Amalthea also lends its name to Capella, diminutive of Capra 
in Latin, meaning ``little female goat''. The name Capra may be used here. 
To finish the naming, Caria, after the region where Cnidos is located, may 
figure as well, after the association with Ctesias.
 
\subsection{Musca}  
\label{Musca}  

Musca, the fly, is one of Plancius' constellations. Although he 
created it as {\it Apis}, the bee, Lacaille renamed it Musca Australis, in
symmetry with then existing northern constellation of Musca Borealis, 
north of Aries. Although a modern constellation, we can draw some 
mythological association. A gadfly appears twice in Greek mythology. 
Jealous Juno sent a gadfly to torment Io, then turned into a cow. 
The other episode is when Bellerophon tried to reach the Olympus riding on Pegasus. 
Jupiter used the same dirty trick, now with almost fatal consequences. Bellerophon 
lost his grip, fell and was saved of certain death by Minerva. 

Yet, the former name Apus serves for a better association, since 
bees play a more prominent role than gadflies in ancient culture. The bee was 
used as an emblem of Potnia, and priestesses of Diana and Minerva 
were referred as bees (Melissa). Melissa is also the name of the 
nymph who discovered honey, and helped Amalthea nourish the infant 
Jupiter. In a later version of the myth, she was given a father, 
Melisseus, or ``honey-man''. In Delphi, the priestesses were also 
called Delphic-bee, according to Pindar. The connexion with Delphi 
allows for a interesting association. Melaina, one of the Naiads, 
is associated with the springs of Delphi, and loved by Apollo. 
Melissa is too common a female name for a planet, but Melaina, 
although not sharing the same etymology, is phonetically similar, as 
thus fit for the name of HD 111232 b.

\subsection{Norma}       
\label{Norma}       

Norma is one of Lacaille's constellations, and as 
such, has no mythological association. It represents a right angle, 
a rule, or a carpenter's square. Its association being related to 
design, architecture and sciences in general, Minerva is the closest 
association. I suggest one of the three planets in Norma be named 
after Metis, Titanian also associated with wisdom, and mother of Minerva, 
albeit in a most unusual birth. 

The other planets could be named Aegis, after the shield 
of Minerva, made by Vulcan himself; Pallas, or Pallax, after one 
of her most widespread titles, Pallas Athena; and Labrys, after the 
double axe that Vulcan used to open Jupiter's skull and give birth 
to Minerva. 

\subsection{Octans}   
\label{Octans}   

Octans is named after the octant, a navigation instrument. Its significance 
is that it is where the celestial south pole is located. Two planets 
are known in Octans, that I propose be named Auster and Notus, two 
names for the South Wind. 
    
\subsection{Ophiuchus}   
\label{Ophiuchus}   

Ophiuchus or Serpentarius, the snake-holder, represents Asclepius, son of 
Apollo and the mythological founder of medicine and healing. Its five planets 
could be called him and his kin. Vediovis (after his Roman name), 
Yaso, Epione and Meditrina, daughters, and Aratus, his son.

\subsection{Orion}
\label{Orion}

Orion the Hunter figures prominently in the sky, but not much in 
Greek mythology, in the sense that his myth is not as well 
developed as, say, Orpheus, Perseus, or 
Theseus. At the time of the Illiad and the Odyssey, he is already 
dead, and Homer and Hesiod already mention him as a constellation. In 
the Odyssey, Ulysses sees him in the underworld, spending the eternity 
hunting animals. He is mentioned in a few lines in the oldest works 
of Greek mythology, but no great work developed the myth to 
the point of creating a standard. 

In one of the most accepted 
versions of the myth, he is a giant son of Neptune, extremely 
handsome, and an excellent hunter. His favorite prey, as mentioned before, 
is the hare, Lepus. Orion falls in love with Merope (not the pleiad), 
seduces her, and is blinded by her father, Oenopion. An Oracle told 
him that his sight could be restored if he travelled to the east and 
exposed his eyes to the rising sun. He made it to Lemnos, where Vulcan 
provided him with a guide, Cedalion. Guided by Cedalion, he met 
his goal and had his sight restored by the sun-god Helios, and Eos, the Dawn. 

The tale of his death is the most variant point of the myth. In one 
version, he bragged that he could kill any beast. The Earth, horrified, 
then sent a giant scorpion, Scorpius, to sting and kill him. In 
a variant, he befriended Diana, going on many hunters with the mighty 
goddess, much to Apollo's dislike. One day, Apollo saw him swimming, 
and dared Diana on arrow shooting using the distant rock as a target. 
The rock was actually Orion's head, and so Diana accidentally killed him. 
Many other variants exist, and as my intent is not to re-tell the myths 
(especially one with so many variants), but to draw names for planets, I 
will allow myself to narrate just one more. This variant combines both stories by 
having Scorpio chasing Orion, who 
swims aways just to be shot by Diana. The position of the constellations 
in the sky, diametrically opposite, perpetuates the chase. 

Four planets have been found around stars of Orion. I propose
Cedalion, Eos, and Sidde. The latter is an earlier love affair of Orion. 
To establish a connection with other mythologies, I also use Gilgamesh, 
the hero after which the Sumerians named the constellation.

\subsection{Pavo} 
\label{Pavo} 

Pavo, the peacock, is one of Plancius' constellations.
Some mythological association can be drawn since the peacock is the 
bird associated with Juno, and one of the main symbols of the 
mighty queen of the gods. 
In the myth, she took the 
hundred eyes of dead Argus and placed it on the peacock's tail. 

I suggest the planets in Pavo be named after Juno. However, since 
Juno is already the name of one of the 
biggest asteroids, her Greek name, Hera, is more suitable. Cithaeron, the 
mountain where she married Jupiter; Argolia, the place of her adoration; 
Cydippe, a priestess of Hera; and Vesta, her sister, goddess of the hearth, 
should complete the naming of the four planets discovered in Pavo. Vesta 
also goes by her Greek name, Hestia, because she also already names a 
major asteroid.

\subsection{Pegasus}      
\label{Pegasus}      

Pegasus, the winged horse, and his brother Chrysaor both 
sprang from the blood of the severed head of Medusa as it 
fecunded the Earth (or in some variants, the ocean). The horse 
was tamed by Bellerophon, or in a variant, by Minerva, who later 
gave the horse to Bellerophon. Pegasus aided Bellerophon 
in his fight against the Amazons and the Chimera, a monster sibling of 
Cerberus and the Hydra. 

There are 11 planets so far discovered in Pegasus, including the very first, 
51 Pegasus b. This planet was actually already nicknamed Bellerophon, 
after the obvious connection. In another obvious connection, I propose 
the name Chrysaor. Other names may be Anteia, who desired Bellerophon; 
Iobates, her father; and Philonoe, her sister. These may name the 
triple system of HR 8799. Chimera 
is also an obvious choice. The fight with the Amazons also 
inspires the inclusion of Penthesilea and her sister Hypolita, 
even though the former is more associated with the Trojan War, 
and the latter with Hercules' labors. 

The presence of Minerva in the myth allows for yet another interesting 
addition. The other two planets then can be drawn from her myth. 
Parthenos, one of her many titles, and Nike, the goddess of victory, 
who follows Minerva.

\subsection{Perseus}   
\label{Perseus}   

Perseus, one of the great heroes of the Greek, has an 
extensively developed myth where to draw suitable names from. I suggest to 
name the five planets known to date in Perseus after Danae, his mother; 
Eurymedon, his title (Perseus Eurymedon, according to Appollonius' 
Argonautica); Seriphos, the island where Danae and infant Perseus were 
ashore, and Kibisis, the pouch where Perseus' held the severed 
head of Medusa. As I do not envision a stiff naming system where the 
Perseids are only in the constellation of Andromeda, I suggest here 
one of the planets be named Nicippe, after Sthenelus' wife. 

\subsection{Phoenix}      
\label{Phoenix}      

Phoenix is one of Plancius' constellations, named after the 
mythological fire-bird. The mythology, however, is not Greek 
but Phoenician. Closely associated 
with the cult of their sun-gods, Pheonicians and Egyptians knew the Phoenix 
as the Bennu, a bird identified with a stork or a heron. The ancient 
Greeks associated it their own word for phoenix, which means crimson.
They and the Romans subsequently pictured the bird more like a 
peacock or an eagle. According to the Greeks, the phoenix lived in 
Phoenicia next to a well. At dawn, it bathed in the water of the well, 
where the sun-god Helios stopped his chariot in order to listen to its song.

We may name the planets in Phoenix after the horses that pulled the chariot 
of Helios. They have different names according to the sources. 
For Homer, two of them are named Abraxas and Therbeeo. For Eumelus 
they are Eous, Aethiops, Bronte or Tonitrua, and Sterope or Fulgitrua. 
Ovid calls them Pyrius, Eous, Aethon, and Phlegon.

\subsection{Pictor}       
\label{Pictor}       

Pictor, representing a painter's easel, is one of Lacaille's constellations. 
Being art-related, Apollo is the possible mythological association. 

Being related to painting, the portraits made in the Renaissance 
also come to mind. I pick one painting in particular, representing 
a famous episode concerning Apollo, that of the duel of Marsyas. 
Marsyas was a flutist who challenged Apollo to 
a musical duel. Apollo naturally won and, as a punishment, hung 
Marsyas on a tree and skinned him alive. Among the judges was Midas, 
the only one who voted for Marsyas. As a punishment for his clearly 
non-musical ear, Apollo changed his ears into donkey's ears. 
 
Midas is also a most interesting character. He once hosted the satyr 
Silenus, who was also Bacchus' foster father. Grateful, Bacchus promises 
to give Midas whatever he wanted. His famous wish was that everything he 
touched be turned into gold. At first excited by his new gift, he offered 
a feast. He soon realized how unwise his wish was as wine and food turned 
into solid hard gold as he touched them; his family and servants also 
going the same fate as he held them in despair. Begging Bacchus to take 
the cursed gift away, he was told to wash himself in the river
Pactolus. Afterwards, Midas lost his lust for richness and lived a pastoral 
life, worshipping Pan. 

Planets in Pictor may therefore be called Marsyas, Midas, Silenus, and 
Pactolus. Also Sardis, after where the duel of Apollo and Marsyas took 
place; and Asellus, meaning donkey, after Midas' punishment for the 
unwise judgement. Lityerses, Midas' son, completes the naming. 

Midas is a suitable name for the planet of beta Pictoris, that is
still on the make inside its protoplanetary disk. As Midas, planets in 
this phase of their evolution have evolved their own version of Midas' 
touch. In their feeding frenzy, they grow so massive that their gravity  
carves a deep gap centered in its feeding zone, thus halting further 
growth. 

\subsection{Pisces}   
\label{Pisces}   

According to an ancient Syrian legend, Pisces represent Venus and 
her son Cupid (Eros). Greek legend recounts that they leapt into 
the Euprathes in order to escape from Typhon, transforming themselves 
into fishes to swim away from the danger. This association allows for the use 
of what is perhaps the most beautiful myth of Greek mythology, that of 
Eros and Psyche. For the 6 planets found in Pisces, I therefore 
propose the names Porus and Penia (poverty and necessity), parents 
of Eros according to one variant of the myth; naturally, Eros and 
Psyche for the double system of  HD 217107; Hedone, after Eros' 
and Psyche's son; and Zephyrus, the West wind, who carried Psyche to 
Eros' cave.

\subsection{Piscis Austrinus}
\label{Piscis Austrinus}

Piscis Austrinus, the southern fish, is seen in the sky drinking water from
Aquarius' jar. Since Aquarius represents Ganymede, I find a good choice that 
Piscis Austrinus be associated with Troy. It is only natural that Ganymede, 
from Olympus, should keep watching over and providing to his home city, where 
his parents still grieve his absence. The planet of 
Fomalhaut b therefore could be named Illion, whereas the other 
one, HD 216770b, Troad. Further planets found there could be named 
Dardania, and Teucria, for the other names of Troy. 

\subsection{Puppis}
\label{Puppis}

Puppis, the poop deck, with Carina and Velorum, constitute the Argo Navis. 
Of the 9 planets found in Puppis, three circle the star HD 69830. 
It is the only triple system in the Argo, and I cannot resist the 
temptation of calling them after the Fates. Clotho, Lachesis, and 
Atropos, who decide the fate of men. It is a good point to introduce 
names from the Lusiads in the Argo, since Jupiter agreed to their voyage because 
the Fates had so decided. For a sailing epic, Navis is but a natural place. 
Lusus, the mythical founder of Lusitania, and Tagide, after the Tagides, 
nymphs of the Tagus, can be used for two planets. 
Back to the argonauts, I randomly pick Lynceus, Iolaus, and Mopsus. 
Medea, the tragic lover of Jason, completes the planet naming in the constellation. 

\subsection{Pyxis}   
\label{Pyxis}   

Pyxis, the mariner's compass, is one of Lacaille's constellations. Being magnetic 
in nature, I tend to associate the constellation with the region of 
Magnesia. Magnes, descendant of Deucalion, and first king of Magnesia is one 
of the names of choice. Of his descendants, Dictys and Hymenaios may serve as well. 

\subsection{Reticulum}   
\label{Reticulum}   

Reticulum, like Telescopium, is in some ways an example of meta-naming, in 
the sense that it is associated with astronomy itself.  
It refers to the reticle, the piece in front of the telescope cross-haired 
in a ${\tt+}$ shape, that renders stars in photographs their distinct 
``spikes''. 

Being astronomy-related, Urania, the heavenly Muse of Astronomy, 
would be the natural choice. The choice may be considered bad, 
since we already have Uranus as a planet in our own solar system. 
Nevertheless, I stick to this choice. Having Uranus 
and Urania featuring in the sky is as confusing as having Rome and Rumania, 
Turkey and Turkomenistan, Niger and Nigeria, i.e., they are simply two 
names with the same radical, Ouranos, the Greek work for sky. 
Providing Uranus with a similarly sounding 
drone is also a good opportunity to 
alleviate the cacophonia of its name in English.

Other muses may integrate the constellation. The eight other muses, 
Calliope (epic poetry), Clio (history), Euterpe (lyric poetry), 
Melpomene (tragedy), Polyhymnia (choral poetry), Terpsichore (dance), 
and Thalia (comedy), have interesting associations on their own,  
and I use some of the names in Ursa Major. The myth of the Muses evolved 
in time, so that prior to the nine usually recognized as nine muses, 
there were others. Pausanias lists three, Aoide (songs), Melete (practice), 
and Mneme (memory), daughters of Uranus and Plusia. Mneme also 
resonates with Mnemosyne, the Titanian that personifies memory. The 
myth of the Titans would of course 
be a good source of names, but they are almost all already used to 
name the moons of Saturn. I already use at least one of them, Saturn himself,
albeit with his Greek name, to name the only planet found in Horologium. 

I therefore pick the names of Melete, Mneme, and Plusia. 
Other names may draw from the Titans, Uranus, and Apollo. The latter 
due to his association with the Muses.

\subsection{Sagitta} 
\label{Sagitta} 

Sagitta, the arrow, represents the arrow that Hercules used to 
kill Ethon, the eagle that tortured Prometheus. In another variant 
it is the arrow of Cupid, or the arrow with which Apollo killed a
cyclops, or yet an arrow that Chiron (Sagittarius) shoots at Scorpion. 

In any case, Sagitta should be a homage to famous archers of the Greek myths. 
There is only one planet so far discovered in Sagitta. I propose to name it 
Paris since he, with an arrow, succeeded where Hector failed 
with a sword. 

\subsection{Sagittarius}       
\label{Sagittarius}       

Sagittarius, the archer, is usually associated with Chiron, the 
first among the Centaurs. Even though Chiron is already a 
minor body of the solar system, naming one of the planets in 
Sagittarius after him is almost unavoidable. In the most accepted version 
of the myth, he is born of the union between Ixion and the 
cloud goddess Nephele, yet an older version has him being the 
fruit of Cronus and Philyra, the nymph who taught humanity 
how to make paper. Chiron's haunts were on 
Mount Pelion; there he married the nymph Chariclo who bore 
him three daughters, Melanippe (or Arne), Endeis, and Okyrhoe, and 
one son, Carystus.

Chiron's figures prominently in mythology as being the mentor 
of many heroes, such as Asclepius, Ajax, Aeneas, Theseus, Achilles, 
Jason, and Hercules. I pick Ajax, Caeneus, Actaeon, Telamon, Patroclus, 
Aeneas, and Achilles among Chiron's disciples for planets in Sagittarius. 

\subsection{Scorpius}          
\label{Scorpius}          

Scorpius, the scorpion, represents the giant scorpion that killed (or 
according to a variant, chases) Orion. There is at least one more 
connection in mythology where the scorpion appears, and that has to 
do with Helios, the sun god, and his son, Phaeteon, the shining-one. 
Helios once promised Phaeteon anything that he wanted, to which 
the boy asked to drive the sun chariot for one day. Helios reluctantly 
consented, which proved a disaster. Phaeteon was not 
able to maneuver, and flew too close to the Earth, burning half 
of Africa and turning it into a desert. In one version, Jupiter 
fulminated him with a lightning 
before he could do more damage. In another version, he flew too close to 
Scorpius, already a constellation, and was stung and killed by him. 

The planets in Scorpion may then be named after this myth and that of 
Helios. I propose Phaeteon and Clymene, his mother. Clymene was also 
mother of the Heliades, Phaeteon sisters, who wept copiously after his death.
They are Aegiale, Aetheria, Helia, and Dioxippe. Leucothoa and 
Euryphaessa are other love interests of Helios. One of his grandchildren, 
Ialysos, may also figure. Phaeteon, who flies or too close or too 
far, is a suitable name for an eccentric planet. One of the Heliades 
was also known as Merope. Merope is also a star of the 
Pleiades, and I purposevely suggest to include her name among the exoplanets, 
for the same reason Atlas was used in \sect{Bootes}. 

In Scorpius, there also figures a planet around a pulsar, 
PSR B1620-26 b. As explained in \sect{sect:naming}, I propose to 
name these after the damned ones in the Tartarus. That particular 
planet was actually already nicknamed ``Methuselah'', due to its 
antiquity. We may as well name it after one of the first creatures 
thrown into the Tartarus, the Hecatonchires, who were later released 
by Jupiter, and helped the Olympics in the war 
against the Titans. However, the name ``Hecatonchir'' sounds unpleasant 
(still better than Methuselah). I propose to call it Erebus. As Erebus 
is also a primordial deity of the Greek creation myth, it suits for 
the name of the ancient planet.  

\subsection{Sculptor}   
\label{Sculptor}   

Sculptor is one of Lacaille's constellations. Lacaille envisioned it 
as Apparatus Sculptoris, the sculptor's studio. The name was later shortened. 

The way of reverting Lacaille's tradition is obvious. The planets in 
Scultor may be named after several sculptors in the Greek myths. The most 
notable of which is Prometheus, the sculptor per excellence, who 
sculpted mankind from clay and gave them life. I prefer to avoid it 
though, since in the current age we still have people believing in 
this creation myth, albeit in its Semitic version. Besides, Prometheus already 
figures in Aquila. 

Another notable mythical sculptor is Pygmalion, who fell 
in love with his own sculpture (in Ovid's metamorphosis). 
He prays to Venus, who concedes life to the 
statue. Of their union springs a son and a daughter, Paphos and 
Metharme. 

\subsection{Scutum}
\label{Scutum}

Scutum, the shield, contains no known planet-hosting stars as of Oct 2009.

\subsection{Serpens} 
\label{Serpens} 

Serpens, the snake, is the reptile being held by Ophiuchus, the snake holder. 
In mythology, it can associated with many monsters, most notably Python and 
Typhon. With Echdina, Typhon fathered the well-known Cerberus and the 
lesser-known Orthrus, a two-headed dog slain by Hercules in his 
tenth labor.  Typhon is said by Homer to dwell in a cave with 
the she-dragon Drakaina, or Delphyne. I propose the double system around 
HD 168443 be named Typhon and Delphyne. Other planes can be named 
Python, Orthrus, Askalaphos (after Asklepius), and Arima, the mythical 
place where Jupiter slain Typhon, also the place where most of 
these monsters were confined.

Also worth noting is that Python was slain by Apollo, 
so Serpens is yet another constellation that can be associated 
with him. Through Cerberus, Pluto's myth can also be used for further 
discovered there.

\subsection{Sextans} 
\label{Sextans} 

Sextans is one of Hevelius' constellations, named after the 
astronomical sextant. As an instrument and thus fruit of human 
intelligence can be associated with 
Minerva. For its astronomical relevance, with Apollo. Yet I draw from its 
use in navigation, to associate it with navigations myth such as the Odyssey,
the Argonauts or the Lusiad. The more natural would be the Lusiad, since the 
sextant was invented in the Middle Ages. However, I prefer to keep 
the associations loose, and draw names from the Odyssey for the four 
planets known in Sextans. Calypso and Circe, two of the many reasons 
that so delayed Ulysses' return to Ithaca; Telegonus and Nausinous, 
kin of Ulysses and Circe. 
          
\subsection{Taurus}   
\label{Taurus}   

Taurus, the bull, is usually associated with the myth of the abduction 
of Europa. The princess of Phoenicia was raptured by Jupiter, 
in the shape of a bull. Another association 
is with the bull of Crete, the seventh of Hercules' labors. 
The association with Europa leads immediately to her most famous 
brother, Cadmus, who Herotodus credits as bringing the 
Phoenician alphabet to Greece, and hence to Europe (Cilix and 
Phoenix were her other brothers). Cadmus is also the mythical founder 
of Thebes, and has a rich myth where to drawn names from. 
Cadmus is also the consort is Concordia (Harmonia), the goddess 
of harmony. A suggestion from Moore, and I concur, is that 
a double system should be named after Corcordia and Pax, peace, her 
sister, as both often go together. However, Moore chose for that the system of 
Gliese 876, in Aquarius. I find Taurus a more proper
place, given the association with Cadmus. The three are suitable names
for the planets around HD 37124. Cadmus allows for yet another 
interesting association. Semele, the mortal mother of Bacchus, is 
his and Corcordia's daughter.  The association 
with an Olympic is a welcoming addition. Semele, Cilix, Cadmus, 
Pax and Corcordia are my choices for the known planets in Taurus. 

\subsection{Telescopium}
\label{Telescopium}
Telescopium contains no known planet-hosting stars as of Oct 2009.

\subsection{Triangulum}   
\label{Triangulum}   

Triangulum is a constellation with no mythological association. 
It is was listed by Ptomely because its brightest 
stars seem to form a small isosceles triangle. 

Representing geometry, it may be associated with Minerva, due to hers 
being patron of mathematics. Also, Pythagoras' reputation was so vast 
and his life so involved in secrecy (because of the Pythagorean 
brotherhood), that he was thought to be ``born to Zeus-beloved Apollo'', 
sent by the gods to benefit humankind, and so described by one of his 
ancient biographers, Iamblichus. Also, the triangle is a percussion 
instrument, and thus also related to Apollo. 

Another possibility are the Dactyls, who worked in the forge of Vulcan and 
are said to have taught metalworking, mathematics, and the alphabet to humans.
In a version of the myth, they go by the name of Acmon (anvil), 
Damnameneus (hammer), and Celmis (casting), which I use for Triangulum and 
Triangulum Australe. 
     
\subsection{Triangulum Australe}
\label{Triangulum Australe}

See Triangulum (\sect{Triangulum}).

\subsection{Tucana}
\label{Tucana}

Tucana, representing the toucan, a bird of South America, is one of 
Plancius' constellations and has no mythological association. As an animal, it may 
be linked with Mercury or Diana. Taking Mercury, I propose for the 
two planets in Tucana the names 
Cylenne, after the mount where Mercury was born; and Nysa, the mount where 
Mercury took the infant Bacchus to be raised by nymphs.

\subsection{Ursa Major} 
\label{Ursa Major} 

Ursa Major, the great bear, represents Callisto, as mentioned in 
Bootes. I draw loosely on the myth, since a good part of it is 
reserved to Bootis. Callisto is a nymph of Diana, and through 
her Ursa Major gets an Olympic association. Planets in Ursa Major 
therefore can be named after Artemis, Diana's Greek name; Leto, 
the mortal mother of Apollo and Diana; Adonis, since Diana is usually 
implicated in his death; Atalanta, the huntress, Diana's proteg\'e and the 
only woman among the Argonauts; Calydon, after the hunt for the Calydonian 
Boar; 
and Niobe, the tragic queen of Thebes and mother of fourteen kids, who 
looked down on Lato's lower fertility rate, just to have all her kin 
wiped out by an angered pair of deities. An ancient example on the old 
quantity vs. quality debate. 

I deliberately cut short the name-drawing from Diana's myth. My purpose is 
to include some from Apollo, to further stress that the naming convention 
I intend to convey is flexible, based on loose, even far-fetched associations. 
The purpose is to maximize name sources from classic antiquity to a particular constellation. 
Even though Apollo was already summoned to several constellations, he 
is Diana's brother. From his quality as patron of arts, I use the names 
of Calliope, Clio and Euterpe, three of the muses, for further planet names 
in Ursa Major. 

And, last but not least, Ursa Major represents in the north what 
Crux is for the southern hemisphere: an easily recognizable asterism 
that points to the pole. Skiron and Kaikias, northwest and northeast wind, 
complete the naming of planets in Ursa Major. 
      
\subsection{Ursa Minor}     
\label{Ursa Minor}     

Ursa Minor, the little bear, containing the north pole, mirrors the Octans in the southern 
hemisphere, although with much more abundance of bright stars. It may 
represent Arcas - although Bootis may as weel -, a dog, or the garden 
of the Hesperides, the stars being golden apples, or the Hesperides 
themselves. In any case, the connection with North is the most obvious. 
Two planets have been found in Ursa Minor (though one of them is doubtful). 
I propose the names Boreas, after the north wind, and Hiperborea - the 
mythical country where the sun shone 24 hours a day. Being associated 
with north, both Ursa Minor and Ursa Major may eventually use names from 
the Valkyries of Norse mythology, female riders send by Odin to decide who 
would die in battles. It was imagined that light reflecting on their 
armors caused the Northern lights.

\subsection{Vela}
\label{Vela}

Vela, along with Carina and Puppis, constitute the Argo Navis. 
I randomly choose the following argonauts: 
Argus, Peleus, Echion, Idmon, and Palaemon. 

\subsection{Virgo}  
\label{Virgo}  

Virgo, the maiden, can be associated with any prominent goddess. Juno, 
Minerva, Diana, Ceres or Proserpina, have all been connected to the 
constellation. The association with Proserpina is the one 
I find most interesting. Her return from the Underworld to the company 
of her mother marks the beginning of spring. Similarly, Virgo is the 
spring constellation in the northern hemisphere. I choose her myth to 
name the planets in Virgo. 

The double system of HD 102272 could be named after Ceres and Proserpina, 
albeit with their Greek names Demeter and Persephone, to celebrate the 
long awaited reunion of mother and daughter. 70 Virginis b, nicknamed Goldilocks, 
may better go by the name of Cora, Persephone's name prior 
to the abduction. Names for 
other planets could be Nycteus and Alastor, two of the four horses Pluto 
used to abduct Persephone; Eleusis, where the abduction took place, according 
to the Greek myth, or Enna according to the Roman; Hecate, goddess of 
darkness, who followed Persephone into Hades 
(in another variant, one of Demeter's guises as a trinity goddess, 
Persephone herself being the other); 
Rodi, Greek for pomegranate; and Narcissus. The later 
because the flower named after the handsome lad is what distracted 
Persephone and her companions, thus facilitating the abduction. His presence 
also allows for yet another myth to be used in the naming, from which I pick 
the nymph Liriope, his mother; the river-god Cephisus, his father; and 
Ameinias, who in an earlier variant of the myth was the rejected lover 
of Narcissus (Echo is already used in \sect{Capricornus}). 

From Ceres' myth, I suggest the names Erinys (anger), one of 
her titles; and Callichoron, the well where she copiously wept for the 
abduction of Persephone. 

The planets around the pulsar PSR 1257+12, as mentioned in \sect{sect:naming}, 
should 
receive names of the damned in the Tartarus. I propose to name them after 
Sisyphus, Ixion, and Tantalus. As the Underworld is the realm of Persephone, 
these three poor souls are in the right place in the sky.  
        
\subsection{Volans}      
\label{Volans}      

Volans, the flying fish, is one of Plancius' constellations. I draw 
again on the myth of Neptune, and for the planet around HD 76700 
I pick a Nereid, Nesaea. She was one of the Nereids who gathered 
round Thetis in her sympathetic grief for Achilles' loss of Patroclus.

\subsection{Vulpecula}          
\label{Vulpecula}          

Vulpecula, the fox, is one of Hevelius' constellations. 
Although Hevelius did not create it for any mythological association, 
we can therefore 
find one fox in Greek mythology. That is the Teumessian fox, a gigantic 
animal of the kin of Echidna. The Teumessian fox had the particular property 
that it could never be caught. It was chased by Laelaps, the dog that 
caught anything. The contradiction that was set caught the attention of 
Jupiter, who elevated both to the skies, which is why in some 
myths it represents Lepus, the prey of Orion. We may as well conclude that 
it simply underscores the philosopher's taste for paradoxes. 

\section{Summary and concluding remarks}

In this manuscript I suggest names for the 403 extrasolar planet candidates 
known as of October 2009. The suggestion is based on the classical tradition 
of giving names from Roman-Greek mythology to astronomical bodies. 
The association with the myths is in many cases purposively 
loose, to enable more flexibility on naming further planets as they are 
discovered. As said in \sect{subsect:nonroman}, the system does not exclude 
other mythologies, which may be used if a suitable association with the 
constellation can be established.

The system also has some power of prediction. We can say for instance, 
that the next planet discovered in Eridanus could be called 
Minos, after one of the three judges of Hades. Indeed, in a former version 
of this manuscript I had suggested that the next three planets could be called 
Minos, Eachus, and Radhamantus. Then, on Sep 17, the Encyclopedia 
of Extrasolar Planets was updated. One of the new planets was in Eridanus, 
and I promptly added it to Table 1 as Radhamantus. In the update of 
Oct 9, another planet in Eridanus was announced. I added it to Table 1 
as Eachus. Along the course of 
writing this manuscript, I could already experience another small benefit 
of having the planets named. Scrolling the long table up and down, it 
was much easier to remember ``Typhon'' than ``HD 168443 c''

I further stress that the system proposed here does not intend to 
supplant the one in vogue. It is not a change, but an addition. Stars are 
known by many names. Merope is also known as 23 Tau, HD 23480, HIC 17608, 
HR 1156, 2MASS J03461958+2356541, and V971 Tau, to name a few. The current 
naming scheme of assigning minor letters to the names of stars will 
of course be kept for scientific publications, much in the same way that 
we use HD 135742 instead of Zubenelschamali and NGC 4755 instead of Jewel Box.
The proper name is a bonus aimed at popular writings.

One drawback I can think of is that it may lead the public to assume that 
the constellations are somehow physical. In some way, the misconception 
already exists. I recall this anecdote, for which 
I unfortunately cannot find the reference, that a 
piece of popular scientific writing once defined a constellation as 
``a group of stars. Up to date, astronomers have found only 88''. 
On the other hand, we can invert the argument and see it as a 
golden opportunity to fight this misconception. Along with the 
names and their associations, it has to be pointed that 
the constellations are human invention and just a useful 
way of mapping the sky. Naming the planets after the myth of the 
constellations is no more misleading than using stellar names such as 
14 Herculi, 70 Virginis or Upsilon Andromedae.

It should also be pointed that this manuscript was sent to the IAU 
Commission 53 on exoplanets, whose majority still opposes the idea of adding 
names to planets. The strongest concern of the commission had to do with the 
definition of an exoplanet. The Encyclopedia of Extrasolar Planets, 
complete as it is, lists candidate planets, and some of them are not 
confirmed. Therefore, some of the candidates assigned names here may as 
well be just low luminosity stars, brown dwarfs, a stellar spot or, as 
noted by a member of the commission, even a mote of dust in the spectrograph. 
In that case, the name should be withdrawn and re-assigned to other, 
confirmed, planet.

\vspace{1cm}

{\it Acknowledgments.} I thank the reading and 
comments of J. Alves, S. Boscardin, 
A. Johansen, M.-M. Mac Low, A. Moitinho, N. Piskunov, 
H. Rocha-Pinto, S. Soter, and P. Tsalmantza, that 
helped me on identifying the main points of critique. 
Interestingly, Portuguese speakers expressed concern regarding the 
use of the Lusiad. Westerns were concerned with the supposed Eurocentrism 
of mainly using Roman-Greek myths. It is great to see that we 
are living in multicultural times and that instead of trying to 
inflate national prides and highlighting boundaries, we are actually 
trying to distance ourselves from them. I also thank M. Schmitz for 
organizing an unorthodox peer review among the IAU Comission 53 
and playing the role of editor.

\onecolumn
\begin{longtable}{llcccrrl}
\caption[Extrasolar planet names]{Extrasolar planet names} \label{table:names} \\\hline
{\bf Constellation} & {\bf Planet} & {\bf Mass}    & {\bf SMA}  & {\bf Ecc.} & {\bf RA} & {\bf Dec} & {\bf Name}\\
                 &              & {\bf ($M_J$)} & {\bf (AU)} &            &          &           &           \\
\hline
 
\endfirsthead
\multicolumn{8}{c}%
{\tablename\ \thetable{} -- continued from previous page} \\\hline
{\bf Constellation} & {\bf Planet} & {\bf Mass}    & {\bf SMA}  & {\bf Ecc.} & {\bf RA} & {\bf Dec} & {\bf Name}\\
                   &              & {\bf ($M_J$)} & {\bf (AU)} &            &          &           &           \\
\hline
 
\endhead
\hline \multicolumn{8}{r}{{Continued on next page}} \\ \hline
\endfoot
\hline \hline
\endlastfoot
Andromeda & 14 And b & 4.8 & 0.83 & 0.0 & 23 31 17 & +39 14 10 & Perses \\
 & ups And b & 0.69 & 0.059 & 0.029 & 01 36 48 & +41 24 38 & Heleus \\
 & ups And c & 1.98 & 0.83 & 0.254 & 01 36 48 & +41 24 38 & Mestor \\
 & ups And d & 3.95 & 2.51 & 0.242 & 01 36 48 & +41 24 38 & Cynurus \\
 & HAT-P-6 b & 1.057 & 0.05235 & 0.0 & 23 39 06 & +42 27 58 & Sthenelus \\
 & WASP-1 b & 0.89 & 0.0382 & 0.0 & 00 20 40 & +31 59 24 & Alkaios \\\hline
Antlia & HD 93083 b & 0.37 & 0.477 & 0.14 & 10 44 20 & -33 34 37 & Palamedes \\\hline
Apus & HD 131664 b & 18.15 & 3.17 & 0.638 & 15 00 06 & -73 32 07 & Virbius \\\hline
Aquarius & Gliese 876 b & 1.935 & 0.20783 & 0.0249 & 22 53 13 & -14 15 13 & Dardanus \\
 & Gliese 876 c & 0.56 & 0.13 & 0.27 & 22 53 13 & -14 15 13 & Tros \\
 & Gliese 876 d & 0.018 & 0.0208067 & 0.0 & 22 53 13 & -14 15 13 & Ilus \\
 & HD 210277 b & 1.23 & 1.1 & 0.472 & 22 09 29 & -07 32 32 & Themiste \\
 & HD 222582 b & 5.11 & 1.35 & 0.76 & 23 41 51 & -05 59 08 & Assaracus \\
 & Gj 849 b & 0.82 & 2.35 & 0.06 & 22 09 40 & -04 38 27 & Capys \\
 & HD 219449 b & 2.9 & 0.3 & -- & 23 15 53 & -09 05 15 & Aigesta \\
 & WASP-6 b & 0.503 & 0.0421 & 0.054 & -- & -- & Teucrus \\\hline
Aquila & CoRoT-3 b & 21.66 & 0.057 & 0.0 & 19 28 13 & 00 07 19 & Cratos \\
 & HD 179079 b & 0.08 & 0.11 & 0.115 & 19 11 10 & -02 38 18 & Epimetheus \\
 & HD 183263 b & 3.69 & 1.52 & 0.38 & 19 28 24 & +08 21 28 & Elpis \\
 & HD 183263 c & 3.82 & 4.25 & 0.253 & 19 28 24 & +08 21 28 & Pithos \\
 & HD 192263 b & 0.72 & 0.15 & 0.0 & 20 13 59 & -00 52 00 & Prometheus \\
 & HD 192699 b & 2.5 & 1.16 & 0.149 & 20 16 06 & +04 34 5 & Hesione \\
 & ksi Aql b & 2.8 & 0.68 & 0.0 & 19 54 15 & +08 27 41 & Ethon \\
 & VB 10 b & 6.4 & 0.36 & 0.98 & 19 16 58 & +05 09 02 & Elbrus \\
 & CoRoT-6 b & 3.3 & -- & -- & -- & -- & Zelus \\\hline
Ara & HD 154672 b & 5.02 & 0.6 & 0.61 & 17 10 05 & -56 26 57 & Pelasgus \\
 & HD 154857 b & 1.8 & 1.2 & 0.47 & 17 11 15 & -56 40 50 & Phassus \\
 & HD 160691 b & 1.676 & 1.5 & 0.128 & 17 44 08 & -51 50 02 & Nyctimus \\
 & HD 160691 c & 0.03321 & 0.09094 & 0.172 & 17 44 08 & -51 50 02 & Peucetis \\
 & HD 160691 d & 0.5219 & 0.921 & 0.0666 & 17 44 08 & -51 50 02 & Caucon \\
 & HD 160691 e & 1.814 & 5.235 & 0.0985 & 17 44 08 & -51 50 02 & Cynaethus \\
 & GJ 674 b & 0.037 & 0.039 & 0.2 & 17 28 40 & -46 53 43 & Stymphalus \\
 & HD 156411 b & 0.75 & -- & -- & 11 35 27 & -32 32 24 & Melaeneus \\
 & GJ 676A b & 4 & -- & -- & 11 52 53 & -50 17 34 & Eumon \\\hline
Aries & HD 12661 b & 2.3 & 0.83 & 0.35 & 02 04 34 & +25 24 51 & Aeetes \\
 & HD 12661 c & 1.57 & 2.56 & 0.2 & 02 04 34 & +25 24 51 & Chalciope \\
 & HD 20367 b & 1.07 & 1.25 & 0.23 & 03 17 40 & +31 07 37 & Colchis \\
 & HIP 14810 b & 3.88 & 0.0692 & 0.1427 & 03 11 14 & +21 05 50 & Chrysomallos \\
 & HIP 14810 d & 0.57 & 1.89 & 0.173 & 03 11 14 & +21 05 50 & Helle \\
 & HIP 14810 c & 1.28 & 0.545 & 0.164 & 03 11 14 & +21 05 50 & Phrixus \\\hline
Auriga & HAT-P-9 b & 0.78 & 0.053 & 0.0 & 07 20 40 & +37 08 26 & Euthenia \\
 & HD 40979 b & 3.32 & 0.811 & 0.23 & 06 04 29 & +44 15 37 & Aglaia \\
 & HD 43691 b & 2.49 & 0.24 & 0.14 & 06 19 35 & +41 05 32 & Lycia \\
 & HD 45350 b & 1.79 & 1.92 & 0.778 & 06 28 45 & +38 57 46 & Eurynome \\
 & HD 49674 b & 0.115 & 0.058 & 0.23 & 06 51 30 & +40 52 03 & Lemnos \\
 & WASP-12 b & 1.41 & 0.0229 & 0.049 & 06 30 33 & +29 40 20 & Vulcan \\\hline
Bootes & HAT-P-4 b & 0.68 & 0.0446 & 0.0 & 15 19 58 & +36 13 47 & Pramnos \\
 & HD 128311 b & 2.18 & 1.099 & 0.25 & 14 36 00 & +09 44 47 & Bacchus \\
 & HD 128311 c & 3.21 & 1.76 & 0.17 & 14 36 00 & +09 44 47 & Aithra \\
 & HD 132406 b & 5.61 & 1.98 & 0.34 & 14 56 55 & +53 22 56 & Atlas \\
 & tau Boo b & 3.9 & 0.046 & 0.018 & 13 47 17 & +17 27 22 & Arcas \\
 & WASP-14 b & 7.725 & 0.037 & 0.095 & 14 33 06 & +21 53 41 & Arcadia \\\hline
Camelopardalis & HD 104985 b & 6.3 & 0.78 & 0.03 & 12 05 15 & +76 54 20 & Triklaria \\
 & HD 33564 b & 9.1 & 1.1 & 0.34 & 05 22 33 & +79 13 52 & Ephesia \\
 & XO-3 b & 11.79 & 0.0454 & 0.26 & 04 21 53 & +57 49 01 & Egeria \\
 & HD 32518 b & 3.04 & 0.59 & 0.01 & 05 09 37 & +69 38 22 & Opalia \\\hline
Cancri & 55 Cnc f & 0.144 & 0.781 & 0.2 & 08 52 37 & +28 20 02 & Stygne \\
 & 55 Cnc b & 0.824 & 0.115 & 0.014 & 08 52 37 & +28 20 02 & Anthelea \\
 & 55 Cnc c & 0.169 & 0.24 & 0.086 & 08 52 37 & +28 20 02 & Teleia \\
 & 55 Cnc d & 3.835 & 5.77 & 0.025 & 08 52 37 & +28 20 02 & Argive \\
 & 55 Cnc e & 0.034 & 0.038 & 0.07 & 08 52 37 & +28 20 02 & Euippe \\
 & HD 73534 b & 1.15 & 3.15 & 0.046 & 08 39 16 & +12 57 37 & Pirene \\\hline
Canes Venatici & HAT-P-12 b & 0.211 & 0.0384 & 0.0 & 13 57 34 & 43 29 37 & Asterion \\\hline
Canis Major & HD 47536 b & 5 & -- & 0.2 & 06 37 47 & -32 20 23 & Leucomelaena \\
 & HD 45364 b & 0.1872 & 0.6813 & 0.1684 & 06 25 38 & -31 28 51 & Maera \\
 & HD 45364 c & 0.6579 & 0.8972 & 0.0974 & 06 25 38 & -31 28 51 & Dromis \\
 & HD 47186 b & 0.07167 & 0.05 & 0.038 & 06 36 09 & -27 37 20 & Cisseta \\
 & HD 47186 c & 0.35061 & 2.395 & 0.249 & 06 36 09 & -27 37 20 & Lampuris \\
 & HD 47536 c & 7 & -- & -- & 06 37 47 & -32 20 23 & Lycoctonus \\
 & HD 43197 b & 0.6 & -- & -- & 00 55 11 & -47 24 21 & Arctophonus \\\hline
Capricornus & HD 202206 b & 17.4 & 0.83 & 0.435 & 21 14 57 & -20 47 21 & Syrinx \\
 & HD 202206 c & 2.44 & 2.55 & 0.267 & 21 14 57 & -20 47 21 & Echo \\
 & HD 204313 b & 4.05 & 3.082 & 0.131 & 21 28 12 & -21 43 35 & Dryope \\\hline
Carina & HD 65216 b & 1.21 & 1.37 & 0.41 & 07 53 4 & -63 38 50 & Jason \\
 & OGLE-TR-111 b & 0.53 & 0.047 & 0.0 & 10 53 1 & -61 24 20 & Laocoon \\
 & OGLE-TR-113 b & 1.32 & 0.0229 & 0.0 & 10 52 24 & -61 26 48 & Iphitos \\
 & OGLE-TR-132 b & 1.14 & 0.0306 & 0.0 & 10 50 34 & -61 57 25 & Autolycus \\
 & OGLE2-TR-L9 b & 4.5 & -- & -- & 11 07 55 & -61 08 46 & Erginus \\
 & OGLE-TR-182 b & 1.01 & 0.051 & 0.0 & 11 09 19 & -61 05 43 & Euryalus \\
 & OGLE-TR-211 b & 1.03 & 0.051 & 0.0 & 10 40 15 & -62 27 20 & Hylas \\
 & HD 63765 b & 0.69 & -- & -- & 17 18 59 & -34 59 48 & Acastus \\\hline
Cassiopeia & HD 240210 b & 6.9 & 1.33 & 0.15 & 23 10 29 & +57 01 46 & Eulimene \\
 & HD 7924 b & 0.029 & 0.057 & 0.17 & -- & -- & Orithya \\
 & HD 17156 b & 3.212 & 0.1623 & 0.6753 & 02 49 44 & +71 45 12 & Thetis \\\hline
Centaurus & HD 101930 b & 0.3 & 0.302 & 0.11 & 11 43 30 & -58 00 24 & Nephele \\
 & HD 102117 b & 0.172 & 0.1532 & 0.106 & 11 44 50 & -58 42 13 & Eurytion \\
 & HD 114386 b & 0.99 & 1.62 & 0.28 & 13 10 39 & -35 03 17 & Pholus \\
 & HD 114729 b & 0.82 & 2.08 & 0.31 & 13 12 44 & -31 52 24 & Nessus \\
 & HD 117207 b & 2.06 & 3.78 & 0.16 & 13 29 21 & -35 34 15 & Rhoecus \\
 & HD 117618 b & 0.19 & 0.28 & 0.39 & 13 32 25 & -47 16 16 & Hylaeus \\
 & HD 121504 b & 0.89 & 0.32 & 0.13 & 13 57 17 & -56 02 24 & Asbolus \\
 & HD 109749 b & 0.28 & 0.0635 & 0.01 & 12 37 16 & -40 48 43 & Amycus \\
 & HD 103197 b & 0.1 & -- & -- & 14 21 23 & -40 23 38 & Hylonome \\
 & HD 125595 b & 0.045 & -- & -- & 01 06 02 & -22 27 11 & Cyllarus \\\hline
Cepheus & gamma Cephei b & 1.6 & 2.044 & 0.115 & 23 39 20 & +77 37 56 & Dannaus \\\hline
Cetus & 81 Cet b & 5.3 & 2.5 & 0.206 & 02 37 42 & -03 23 46 & Stheno \\
 & BD-17 63 b & 5.1 & 1.34 & 0.54 & 00 28 34 & -16 13 35 & Thoosa \\
 & HD 11506 c & 0.82 & 0.639 & 0.42 & 01 52 51 & -19 30 25 & Charybdis \\
 & HD 11964 b & 0.11 & 0.229 & 0.15 & 01 57 09 & -10 14 32 & Scylla \\
 & HD 16141 b & 0.23 & 0.35 & 0.21 & 02 35 19 & -03 33 38 & Odysseus \\
 & HD 19994 b & 2 & 1.3 & 0.2 & 03 12 46 & -01 11 45 & Telemachus \\
 & HD 224693 b & 0.71 & 0.233 & 0.05 & 23 59 54 & -22 25 41 & Euryale \\
 & HD 2638 b & 0.48 & 0.044 & 0.0 & 00 29 59 & -05 45 50 & Phorcys \\
 & HD 11506 b & 3.44 & 2.43 & 0.22 & 01 52 51 & -19 30 25 & Echidna \\
 & HD 11964 c & 0.7 & 3.167 & 0.3 & 01 57 09 & -10 14 32 & Deino \\
 & HD 5319 b & 1.94 & 1.75 & 0.12 & 00 55 01 & +00 47 22 & Polyphemus \\
 & HD 6718 b & 1.65 & -- & -- & 01 33 17 & -38 14 42 & Callidice \\
 & HIP 5158 b & 1.3 & -- & -- & 14 20 54 & -17 28 53 & Polypoites \\\hline
Chamaeleon & HD 63454 b & 0.38 & 0.036 & 0.0 & 07 39 21 & -78 16 44 & Naiad \\
 & CT Cha b & 17 & 440 & -- & 11 04 09 & -76 27 19 & Nereus \\\hline
Columba & HD 43848 b & 25 & 3.4 & 0.69 & 06 16 31 & -40 31 55 & Peristera \\\hline
Coma Berenices & HD 108874 b & 1.36 & 1.051 & 0.07 & 12 30 26 & +22 52 47 & Ariadne \\
 & HD 108874 c & 1.018 & 2.68 & 0.25 & 12 30 26 & +22 52 47 & Theseus \\
 & HD 114762 b & 11.02 & 0.3 & 0.34 & 13 12 19 & +17 31 01 & Naxos \\\hline
Corona Borealis & kappa CrB b & 1.8 & 2.7 & 0.19 & 15 51 14 & +35 39 27 & Euanthes \\
 & rho CrB b & 1.04 & 0.22 & 0.04 & 16 01 03 & +33 18 51 & Staphylus \\
 & XO-1 b & 0.9 & 0.0488 & 0.0 & 16 02 12 & +28 10 11 & Latramys \\\hline
Corvus & HD 104067 b & 0.16 & -- & -- & 01 07 49 & -08 14 01 & Coronis \\\hline
Crater & BD-10 3166 b & 0.48 & 0.046 & 0.07 & 10 58 28 & -10 46 13 & Alexiares \\
 & HD 96167 b & 0.68 & 1.3 & 0.71 & 11 05 15 & -10 17 29 & Aniketos \\\hline
Crux & NGC 4349 No 127 b & 19.8 & 2.38 & 0.19 & 12 24 08 & -61 52 18 & Livas \\
 & HD 108147 b & 0.4 & 0.104 & 0.498 & 12 25 46 & -64 01 19 & Apeliotes \\\hline
Cygnus & HAT-P-7 b & 1.8 & 0.0379 & 0.0 & 19 28 59 & +47 58 10 & Leda \\
 & HD 185269 b & 0.94 & 0.077 & 0.3 & 19 37 12 & +28 30 00 & Eurypylus \\
 & HD 187123 b & 0.52 & 0.042 & 0.03 & 19 46 57 & +34 25 15 & Timandra \\
 & HD 190360 b & 1.502 & 3.92 & 0.36 & 20 03 37 & +29 53 48 & Eurythemis \\
 & HD 190360 c & 0.057 & 0.128 & 0.01 & 20 03 37 & +29 53 48 & Thestius \\
 & 16 Cyg B b & 1.68 & 1.68 & 0.689 & 19 41 51 & +50 31 03 & Althaea \\
 & HAT-P-11 b & 0.081 & 0.053 & 0.198 & 19 50 50 & +48 04 51 & Iphicles \\
 & HD 187123 c & 1.99 & 4.89 & 0.252 & 19 46 57 & +34 25 15 & Echemus \\\hline
Delphinus & 18 Del b & 10.3 & 2.6 & 0.08 & 20 58 26 & +10 50 21 & Delphi \\
 & HD 195019 b & 3.7 & 0.1388 & 0.014 & 20 28 17 & +18 46 12 & Apollo \\
 & HD 196885 b & 2.58 & 2.37 & 0.462 & 20 39 51 & +11 14 58 & Melikertes \\
 & WASP-2 b & 0.914 & 0.03138 & 0.0 & 20 30 54 & +06 25 46 & Portunes \\\hline
Dorado & HD 30177 b & 9.17 & 3.86 & 0.3 & 04 41 54 & -58 01 14 & Tyro \\
 & HD 28254 b & 1.16 & -- & -- & 05 23 22 & -02 16 39 & Enipeus \\\hline
Draco & 42 Dra b & 3.88 & 1.19 & 0.38 & 18 25 59 & +65 33 49 & Ladon \\
 & HD 139357 b & 9.76 & 2.36 & 0.1 & 15 35 16 & +53 55 20 & Aegle \\
 & HD 167042 b & 1.6 & 1.3 & 0.03 & 18 10 32 & +54 17 12 & Eryteis \\
 & HIP 75458 b & 8.82 & 1.275 & 0.7124 & 15 24 55 & +58 57 57 & Lipara \\
 & TrES-2  & 1.199 & 0.03556 & 0.0 & 19 07 14 & +49 18 59 & Chrysothemis \\\hline
Eridanus & eps Eridani b & 1.55 & 3.39 & 0.702 & 03 32 55 & -09 27 29 & Styx \\
 & Gl 86 b & 4.01 & 0.11 & 0.046 & 02 10 14 & -50 50 00 & Aqueron \\
 & HD 10647 b & 0.91 & 2.1 & 0.18 & 01 42 29 & -53 44 27 & Cocytus \\
 & HD 28185 b & 5.7 & 1.03 & 0.07 & 04 26 26 & -10 33 02 & Phlegethon \\
 & HD 30562 b & 1.29 & 2.3 & 0.76 & 04 48 36 & -05 40 27 & Radhamantus \\
 & HIP 12961 b & 0.47 & -- & -- & 09 50 02 & -49 47 25 & Eachus \\\hline
Fornax & HD 16417 b & 0.069 & 0.14 & 0.2 & 02 36 59 & -34 34 41 & Etna \\
 & HD 20782 b & 1.9 & 1.381 & 0.97 & 03 20 03 & -28 51 14 & Lipari \\
 & HD 20868 b & 1.99 & 0.947 & 0.75 & 03 20 43 & -33 43 48 & Milos \\\hline
Gemini & HD 50554 b & 4.9 & 2.38 & 0.42 & 06 54 42 & +24 14 44 & Romulus \\
 & HD 62509 b & 2.9 & 1.69 & 0.02 & 07 45 18 & +28 01 34 & Remus \\
 & HD 59686 b & 5.25 & 0.911 & 0.0 & 07 31 48 & +17 05 09 & Lupa \\\hline
Grus & HD 208487 b & 0.45 & 0.49 & 0.32 & 21 57 19 & -37 45 49 & Megarus \\
 & HD 213240 b & 4.5 & 2.03 & 0.45 & 22 31 00 & -49 25 59 & Deucalion \\
 & HD 216435 b & 1.49 & 2.7 & 0.34 & 22 53 37 & -48 35 53 & Pyrrha \\
 & GJ 832 b & 0.64 & 3.4 & 0.12 & 21 33 34 & -49 00 32 & Gerania \\\hline
Hercules & 14 Her b & 4.64 & 2.77 & 0.369 & 16 10 23 & +43 49 18 & Cerenytis \\
 & HAT-P-2 b & 9.09 & 0.06878 & 0.5171 & 16 20 36 & +41 02 53 & Erymanthus \\
 & HD 149026 b & 0.359 & 0.04313 & 0.0 & 16 30 29 & +38 20 50 & Augean \\
 & HD 154345 b & 0.947 & 4.19 & 0.044 & 17 02 36 & +47 04 55 & Alpheus \\
 & HD 155358 b & 0.89 & 0.628 & 0.112 & 17 09 35 & +33 21 21 & Peneus \\
 & HD 155358 c & 0.504 & 1.224 & 0.176 & 17 09 35 & +33 21 21 & Stymphalia \\
 & HD 164922 b & 0.36 & 2.11 & 0.05 & 18 02 30 & +26 18 46 & Diomedes \\
 & TrES-3  & 1.92 & 0.0226 & 0.0 & 17 52 07 & +37 32 46 & Geryon \\
 & TrES-4  & 0.919 & 0.05091 & 0.0 & 17 53 13 & 37 12 42 & Cerberus \\\hline
Horologium & HR 810 b & 1.94 & 0.91 & 0.24 & 02 42 31 & -50 48 12 & Cronus \\\hline
Hydra & 2M1207 b & 4 & 46 & -- & 12 07 33 & -39 32 54 & Lerna \\
 & HD 122430 b & 3.71 & 1.02 & 0.68 & 14 02 22 & -27 25 47 & Adiante \\
 & HD 70573 b & 6.1 & 1.76 & 0.4 & 08 22 50 & +01 51 34 & Amphitrite \\
 & HD 72659 b & 2.96 & 4.16 & 0.2 & 08 34 03 & -01 34 05 & Galatea \\
 & HD 74156 b & 1.88 & 0.294 & 0.64 & 08 42 25 & +04 34 41 & Amymone \\
 & HD 74156 c & 8.03 & 3.85 & 0.43 & 08 42 25 & +04 34 41 & Thaleia \\
 & HD 74156 d & 0.396 & 1.01 & 0.25 & 08 42 25 & +04 34 41 & Pasithea \\
 & HD 82943 b & 1.75 & 1.19 & 0.219 & 09 34 50 & -12 07 46 & Nausithoe \\
 & HD 82943 c & 2.01 & 0.746 & 0.359 & 09 34 50 & -12 07 46 & Menippe \\
 & WASP-15 b & 0.542 & 0.0499 & 0.0 & 13 55 43 & -32 09 35 & Asia \\
 & HD 86264 b & 7 & 2.86 & 0.7 & 09 56 58 & -15 53 42 & Hyperippe \\
 & GJ 433 b & 0.019 & -- & -- & 06 13 36 & -29 53 50 & Spio \\
 & HD 90156 b & 0.055 & -- & -- & 22 46 37 & -56 35 58 & Ianira \\\hline
Hydrus & GJ 3021 b & 3.32 & 0.49 & 0.505 & 00 16 12 & -79 51 04 & Delos \\
 & HD 11977 b & 6.54 & 1.93 & 0.4 & 01 54 56 & -67 38 50 & Ortygia \\\hline
Indus & HD 216437 b & 2.1 & 2.7 & 0.34 & 22 54 39 & -70 04 25 & Hesper \\\hline
Lacerta & HAT-P-1 b & 0.524 & 0.0553 & 0.067 & 22 57 47 & +38 40 30 & Abas \\\hline
Leo & BD20 2457 b & 21.42 & 1.45 & 0.15 & 10 16 45 & +19 53 29 & Omphale \\
 & BD20 2457 c & 12.47 & 2.01 & 0.18 & 10 16 45 & +19 53 29 & Lamus \\
 & GJ 436 b & 0.072 & 0.02872 & 0.15 & 11 42 11 & +26 42 23 & Nemea \\
 & HD 100777 b & 1.16 & 1.03 & 0.36 & 11 35 52 & -04 45 21 & Elissos \\
 & HD 81040 b & 6.86 & 1.94 & 0.526 & 09 23 47 & +20 21 52 & Iraklion \\
 & HD 88133 b & 0.22 & 0.047 & 0.11 & 10 10 07 & +18 11 12 & Polynices \\
 & HD 89307 b & 1.78 & 3.27 & 0.241 & 10 18 21 & +12 37 15 & Lycurgus \\
 & HD 99109 b & 0.502 & 1.105 & 0.09 & 11 24 17 & -01 31 44 & Cleonae \\
 & HD 99492 b & 0.109 & 0.1232 & 0.254 & 11 26 46 & +03 00 22 & Tmolus \\\hline
Leo Minor & HD 87883 b & 1.78 & 3.6 & 0.53 & 10 08 43 & +34 14 32 & Archemoros \\\hline
Lepus & HD 33283 b & 0.33 & 0.168 & 0.48 & 05 08 01 & -26 47 50 & Epimelius \\\hline
Libra & Gl 581 c & 0.01686 & 0.07 & 0.17 & 15 19 26 & -07 43 20 & Eirene \\
 & Gl 581 d & 0.02231 & 0.22 & 0.38 & 15 19 26 & -07 43 20 & Dike \\
 & Gl 581 b & 0.0492 & 0.041 & 0.0 & 15 19 26 & -07 43 20 & Themis \\
 & Gl 581 e & 0.006104 & 0.03 & 0.0 & 15 19 26 & -07 43 20 & Eunomia \\
 & HD 134987 b & 1.58 & 0.78 & 0.24 & 15 13 28 & -25 18 33 & Astraea \\
 & HD 141937 b & 9.7 & 1.52 & 0.41 & 15 52 17 & -18 26 09 & Dysis \\\hline
Lupus & Lupus-TR-3 b & 0.81 & 0.0464 & 0.0 & 15 30 19 & -42 58 46 & Lycaon \\
 & GQ Lup b & 21.5 & 103 & -- & 15 49 12 & -35 39 03 & Therion \\
 & HIP 70849 b & > & 5 & -- & -- & 14 20 54 & Fenris \\\hline
Lynx & 6 Lyn b & 2.4 & 2.2 & 0.134 & 06 30 47 & +58 09 46 & Scythia \\
 & XO-4 b & 1.72 & 0.0555 & 0.0 & 07 21 33 & +58 16 05 & Metanira \\
 & XO-5 b & 1.077 & 0.0487 & 0.0 & 07 46 52 & +39 05 41 & Lyncus \\
 & HD 75898 b & 1.48 & 0.737 & 0.35 & 08 53 51 & +33 03 25 & Celeus \\
 & WASP-13 b & 0.46 & 0.0527 & 0.0 & -- & -- & Doso \\
 & XO-2 b & 0.57 & 0.0369 & 0.0 & 07 48 07 & +50 13 33 & Demophon \\\hline
Lyra & HD 173416 b & 2.7 & 1.16 & 0.21 & 18 43 36 & +36 33 24 & Orpheus \\
 & HD 177830 b & 1.28 & 1 & 0.43 & 19 05 20 & +25 55 14 & Eurydice \\
 & HD 178911 B b & 6.292 & 0.32 & 0.1243 & 19 09 03 & +34 35 59 & Siren \\
 & TrES-1  & 0.61 & 0.0393 & 0.0 & 19 04 09 & +36 37 57 & Hermes \\
 & HAT-P-5 b & 1.06 & 0.04075 & 0.0 & 18 17 37 & +36 37 18 & Petasus \\
 & WASP-3 b & 1.76 & 0.0317 & 0.0 & 18 34 32 & +35 39 42 & Alipes \\\hline
Mensa & HD 39091 b & 10.35 & 3.29 & 0.62 & 05 37 09 & -80 28 08 & Adamastor \\\hline
Microscopium & WASP-7 b & 0.96 & 0.0618 & 0.0 & 20 44 10 & -39 13 31 & Talos \\\hline
Monoceros & HD 45652 b & 0.47 & 0.23 & 0.38 & 06 29 13 & +10 56 02 & Capra \\
 & HD 46375 b & 0.249 & 0.041 & 0.04 & 06 33 12 & +05 27 46 & Dejanira \\
 & HD 52265 b & 1.13 & 0.49 & 0.29 & 07 00 18 & -05 22 01 & Adamanthea \\
 & HD 66428 b & 2.82 & 3.18 & 0.465 & 08 03 28 & -01 09 45 & Caria \\
 & CoRoT-4 b & 0.72 & 0.09 & 0.0 & 06 48 47 & -00 40 22 & Cornucopia \\
 & CoRoT-7 b & 0.0151 & 0.0172 & 0.0 & 06 43 49 & -01 03 46 & Icarus \\
 & CoRoT-7 c & 0.0264 & 0.046 & 0.0 & 06 43 49 & -01 03 46 & Daedalus \\
 & CoRoT-1 b & 1.03 & 0.0254 & 0.0 & 06 48 19 & -03 06 08 & Achelous \\
 & CoRoT-5 b & 0.467 & 0.04947 & 0.09 & 06 45 07 & 00 48 55 & Cybele \\
 & HD 44219 b & 0.58 & -- & -- & 01 23 37 & -41 16 11 & Dexamenus \\\hline
Musca & HD 111232 b & 6.8 & 1.97 & 0.2 & 12 48 51 & -68 25 30 & Melaina \\\hline
Norma & HD 142415 b & 1.62 & 1.05 & 0.5 & 15 57 40 & -60 12 00 & Metis \\
 & HD 143361 b & 3.12 & 2 & 0.15 & 16 01 50 & -44 26 04 & Aegis \\
 & HD 330075 b & 0.76 & 0.043 & 0.0 & 15 49 37 & -49 57 48 & Pallas \\
 & HD 148156 b & 0.91 & -- & -- & 02 46 43 & -23 05 12 & Labrys \\\hline
Octans & HD 142022 A b & 4.4 & 2.8 & 0.57 & 16 10 15 & -84 13 53 & Auster \\
 & HD 212301 b & 0.45 & 0.036 & 0.0 & 22 27 30 & -77 43 04 & Notus \\\hline
Ophiuchus & HD 148427 b & 0.96 & 0.93 & 0.16 & 16 28 28 & -13 23 59 & Vediovis \\
 & HD 156846 b & 10.45 & 0.99 & 0.8472 & 17 20 34 & -19 20 01 & Yaso \\
 & HD 170469 b & 0.67 & 2.24 & 0.11 & 18 29 11 & +11 41 44 & Epione \\
 & HD 171028 b & 1.83 & 1.29 & 0.61 & 18 32 15 & +06 56 45 & Meditrina \\
 & HD 149143 b & 1.33 & 0.053 & 0.016 & 16 32 51 & +02 05 05 & Aratus \\\hline
Orion & HD 37605 b & 2.3 & 0.25 & 0.677 & 05 40 01 & +06 03 38 & Cedalion \\
 & HD 38529 b & 0.78 & 0.129 & 0.29 & 05 46 34 & +01 10 05 & Eos \\
 & HD 38529 c & 12.7 & 3.68 & 0.36 & 05 46 34 & +01 10 05 & Sidde \\
 & HD 290327 b & 2.54 & -- & -- & 20 11 31 & -64 37 14 & Gilgamesh \\\hline
Pavo & HD 181433 d & 0.54 & 3 & 0.48 & 19 25 10 & -66 28 08 & Cithaeron \\
 & HD 196050 b & 3 & 2.5 & 0.28 & 20 37 51 & -60 38 04 & Hera \\
 & HD 181433 b & 0.0238 & 0.08 & 0.396 & 19 25 10 & -66 28 08 & Hestia \\
 & HD 181433 c & 0.64 & 1.76 & 0.28 & 19 25 10 & -66 28 08 & Argolia \\
 & HD 190984 b & 3.1 & -- & -- & 14 29 19 & -46 27 50 & Cydippe \\\hline
Pegasus & 51 Peg b & 0.468 & 0.052 & 0.0 & 22 57 27 & +20 46 07 & Bellerophon \\
 & BD14 4559 b & 1.47 & 0.777 & 0.29 & 21 13 36 & +14 41 22 & Chimera \\
 & HD 209458 b & 0.685 & 0.04707 & 0.07 & 22 03 10 & +18 53 04 & Minerva \\
 & HD 210702 b & 2 & 1.17 & 0.152 & 22 11 51 & +16 02 2 & Hypolita \\
 & HD 219828 b & 0.066 & 0.052 & 0.0 & 23 18 47 & +18 38 45 & Penthesilea \\
 & HR 8799 b & 7 & 68 & -- & 23 07 29 & +21 08 03 & Philonoe \\
 & HR 8799 c & 10 & 38 & -- & 23 07 29 & +21 08 03 & Iobates \\
 & HR 8799 d & 10 & 24 & -- & 23 07 29 & +21 08 03 & Anteia \\
 & V391 Peg b & 3.2 & 1.7 & 0.0 & 22 04 12 & +26 25 08 & Chrysaor \\
 & HAT-P-8 b & 1.52 & 0.0487 & 0.0 & 22 52 10 & +35 26 50 & Nike \\
 & WASP-10 b & 3.06 & 0.0371 & 0.057 & 23 15 58 & +31 27 46 & Parthenos \\\hline
Perseus & HD 16175 b & 4.4 & 2.1 & 0.59 & 02 37 02 & +42 03 45 & Danae \\
 & HD 16760 b & 14.3 & 1.13 & 0.067 & 02 42 21 & +38 37 07 & Nicippe \\
 & HD 17092 b & 4.6 & 1.29 & 0.166 & 02 46 22 & +49 39 11 & Eurymedon \\
 & HD 23596 b & 7.19 & 2.72 & 0.314 & 03 48 00 & +40 31 50 & Seriphos \\
 & WASP-11/HAT-P-10 b & 0.46 & 0.0439 & 0.0 & 03 09 29 & 30 40 25 & Kibisis \\\hline
Phoenix & HD 142 b & 1 & 0.98 & 0.38 & 00 06 19 & -49 04 30 & Pyrius \\
 & HD 2039 b & 4.85 & 2.19 & 0.68 & 00 24 20 & -56 39 00 & Phlegon \\
 & HD 6434 b & 0.48 & 0.15 & 0.3 & 01 04 40 & -39 29 17 & Bronte \\
 & WASP-18 b & 10.3 & 0.02026 & 0.0092 & 01 37 25 & -45 40 40 & Fulgitrua \\
 & WASP-5 b & 1.637 & 0.02729 & 0.0 & 23 57 24 & -41 16 38 & Tonitrua \\
 & WASP-4 b & 1.1215 & 0.023 & 0.0 & 23 34 15 & -42 03 41 & Abraxas \\
 & HD 5388 b & 1.96 & -- & -- & 19 22 53 & -32 55 09 & Aethon \\
 & HD 8535 b & 0.63 & -- & -- & 10 00 48 & -09 31 00 & Therbeeo \\\hline
Pictor & AB Pic b & 13.5 & 275 & -- & 06 19 12 & -58 03 15 & Silenus \\
 & beta Pic b & 8 & 8 & -- & 05 47 17 & -51 03 59 & Midas \\
 & HD 40307 b & 0.0132 & 0.047 & 0.0 & 05 54 04 & -60 01 24 & Asellus \\
 & HD 40307 c & 0.0216 & 0.081 & 0.0 & 05 54 04 & -60 01 24 & Marsyas \\
 & HD 40307 d & 0.0288 & 0.134 & 0.0 & 05 54 04 & -60 01 24 & Sardis \\
 & HD 41004 A b & 2.3 & 1.31 & 0.39 & 05 59 49 & -48 14 22 & Pactolus \\
 & HD 41004 B b & 18.4 & 0.0177 & 0.081 & 05 59 50 & -48 14 23 & Lityerses \\\hline
Pisces & HD 10697 b & 6.12 & 2.13 & 0.11 & 01 44 55 & +20 04 59 & Hedone \\
 & HD 217107 b & 1.33 & 0.073 & 0.132 & 22 58 15 & -02 23 42 & Eros \\
 & HD 217107 c & 2.49 & 5.27 & 0.517 & 22 58 15 & -02 23 42 & Psyche \\
 & HD 3651 b & 0.2 & 0.284 & 0.63 & 00 39 21 & +21 15 01 & Porus \\
 & HD 4203 b & 1.65 & 1.09 & 0.46 & 00 44 41 & +20 26 56 & Penia \\
 & HD 8574 b & 2.23 & 0.76 & 0.4 & 01 25 12 & +28 34 00 & Zephyrus \\\hline
Piscis Austrinus & Fomalhaut b & 3 & 115 & 0.11 & 22 57 39 & -29 37 20 & Illion \\
 & HD 216770 b & 0.65 & 0.46 & 0.37 & 22 55 53 & -26 39 31 & Troad \\\hline
Puppis & HD 48265 b & 1.16 & 1.51 & 0.18 & 06 40 02 & -48 32 31 & Medea \\
 & HD 50499 b & 1.71 & 3.86 & 0.23 & 06 52 02 & -33 54 56 & Lynceus \\
 & HD 60532 b & 3.15 & 0.77 & 0.278 & 07 34 03 & -22 17 46 & Iolaus \\
 & HD 60532 c & 7.46 & 1.58 & 0.038 & 07 34 03 & -22 17 46 & Mopsus \\
 & HD 69830 b & 0.033 & 0.0785 & 0.1 & 08 18 23 & -12 37 55 & Lachesis \\
 & HD 69830 c & 0.038 & 0.186 & 0.13 & 08 18 23 & -12 37 55 & Atropos \\
 & HD 69830 d & 0.058 & 0.63 & 0.07 & 08 18 23 & -12 37 55 & Clotho \\
 & HD 70642 b & 2 & 3.3 & 0.1 & 08 21 28 & -39 42 19 & Lusus \\
 & NGC 2423 3 b & 10.6 & 2.1 & 0.21 & 07 37 09 & -13 54 24 & Tagide \\\hline
Pyxis & HD 73256 b & 1.87 & 0.037 & 0.03 & 08 36 23 & -30 02 15 & Magnes \\
 & HD 73267 b & 3.06 & 2.198 & 0.256 & 08 36 18 & -34 27 36 & Dictys \\
 & GJ 317 b & 1.2 & 0.95 & 0.193 & 08 40 59 & -23 27 23 & Hymenaios \\\hline
Reticulum & HD 23079 b & 2.61 & 1.65 & 0.1 & 03 39 43 & -52 54 57 & Urania \\
 & HD 23127 b & 1.5 & 2.4 & 0.44 & 03 39 24 & -60 04 40 & Melete \\
 & HD 27442 b & 1.28 & 1.18 & 0.07 & 04 16 29 & -59 18 07 & Mneme \\
 & HD 27894 b & 0.62 & 0.122 & 0.049 & 04 20 47 & -59 24 39 & Plusia \\\hline
Sagitta & HD 231701 b & 1.78 & 0.556 & 0.1 & 19 32 04 & +16 28 27 & Paris \\\hline
Sagittarius & HD 169830 b & 2.88 & 0.81 & 0.31 & 18 27 49 & -29 49 00 & Chiron \\
 & HD 169830 c & 4.04 & 3.6 & 0.33 & 18 27 49 & -29 49 00 & Philyra \\
 & HD 179949 b & 0.95 & 0.045 & 0.022 & 19 15 33 & -24 10 45 & Nauplius \\
 & HD 187085 b & 0.75 & 2.05 & 0.47 & 19 49 33 & -37 46 50 & Pelion \\
 & HD 190647 b & 1.9 & 2.07 & 0.18 & 20 07 20 & -35 32 19 & Chariclo \\
 & MOA-2007-BLG-192-L b & 0.01 & 0.62 & -- & 18 08 04 & -27 09 00 & Endeis \\
 & OGLE-06-109L b & 0.71 & 2.3 & -- & 17 52 35 & -30 05 16 & Melanippe \\
 & OGLE-06-109L c & 0.27 & 4.6 & 0.11 & 17 52 35 & -30 05 16 & Okyrhoe \\
 & OGLE-TR-10 b & 0.63 & 0.04162 & 0.0 & 17 51 28 & -29 52 34 & Carystos \\
 & OGLE-TR-56 b & 1.29 & 0.0225 & 0.0 & 17 56 35 & -29 32 21 & Ajax \\
 & OGLE235-MOA53 b & 2.6 & 5.1 & -- & 18 05 16 & -28 53 42 & Caeneus \\
 & SWEEPS-04  & 3.8 & 0.055 & -- & 17 58 54 & -29 11 21 & Actaeon \\
 & SWEEPS-11  & 9.7 & 0.03 & -- & 17 59 03 & -29 11 54 & Telamon \\
 & MOA-2007-BLG-400-L b & 0.9 & 0.85 & -- & 18 09 42 & -29 13 27 & Achilles \\
 & OGLE-05-169L b & 0.04 & 2.8 & -- & 18 06 05 & -30 43 57 & Arne \\
 & HD 171238 b & 2.6 & 2.54 & 0.4 & 18 34 44 & -28 04 20 & Patroclus \\
 & HD 181720 b & 0.37 & -- & -- & 10 23 55 & -29 38 44 & Aeneas \\\hline
Scorpius & HD 145377 b & 5.76 & 0.45 & 0.307 & 16 11 36 & -27 04 41 & Aetheria \\
 & HD 147513 b & 1 & 1.26 & 0.52 & 16 24 01 & -39 11 34 & Clymene \\
 & HD 153950 b & 2.73 & 1.28 & 0.34 & 17 04 31 & -43 18 35 & Aegiale \\
 & HD 159868 b & 1.7 & 2 & 0.69 & 17 38 60 & -43 08 44 & Phaeteon \\
 & HD 162020 b & 13.75 & 0.072 & 0.277 & 17 50 38 & -40 19 06 & Helia \\
 & OGLE-05-071L b & 3.5 & 3.6 & -- & 17 50 09 & -34 40 23 & Dioxippe \\
 & OGLE-05-390L b & 0.017 & 2.1 & -- & 17 54 19 & -30 22 38 & Leucothoa \\
 & PSR B1620-26 b & 2.5 & 23 & -- & 16 23 38 & -26 31 53 & Erebus \\
 & MOA-2008-BLG-310-L b & 0.23 & 1.25 & -- & 17 54 14 & -34 46 41 & Ialysos \\
 & WASP-17 b & 0.49 & 0.051 & 0.129 & 15 59 51 & -28 03 42 & Euryphaessa \\
 & GJ 667C b & 0.018 & -- & -- & 07 47 50 & -54 15 51 & Merope \\\hline
Sculptor & HD 4113 b & 1.56 & 1.28 & 0.903 & 00 43 13 & -37 58 57 & Pygmalion \\
 & HD 4208 b & 0.8 & 1.67 & 0.05 & 00 44 26 & -26 30 56 & Metharme \\
 & HD 9578 b & 0.62 & -- & -- & 22 46 37 & -56 35 58 & Paphos \\\hline
Serpens & CoRoT-2 b & 3.31 & 0.0281 & 0.0 & 19 27 07 & +01 23 02 & Arima \\
 & HD 136118 b & 11.9 & 2.3 & 0.37 & 15 18 55 & -01 35 32 & Python \\
 & HD 168443 b & 8.02 & 0.3 & 0.5286 & 18 20 04 & -09 35 34 & Delphyne \\
 & HD 168443 c & 18.1 & 2.91 & 0.2125 & 18 20 04 & -09 35 34 & Typhon \\
 & HD 168746 b & 0.23 & 0.065 & 0.081 & 18 21 49 & -11 55 21 & Askalaphos \\
 & HD 175541 b & 0.61 & 1.03 & 0.33 & 18 55 41 & +04 15 55 & Orthrus \\\hline
Sextans & HD 86081 b & 1.5 & 0.039 & 0.008 & 09 56 06 & -03 48 30 & Calypso \\
 & HD 92788 b & 3.86 & 0.97 & 0.27 & 10 42 48 & -02 11 01 & Circe \\
 & BD-082823 b & 0.045 & -- & -- & 17 19 51 & -48 32 58 & Telegonus \\
 & BD-082823 c & 0.33 & -- & -- & 17 30 11 & -51 38 13 & Nausinous \\\hline
Taurus & eps Tau b & 7.6 & 1.93 & 0.151 & 04 28 37 & +19 10 50 & Semele \\
 & HD 37124 b & 0.61 & 0.53 & 0.055 & 05 37 02 & +20 43 50 & Cadmus \\
 & HD 37124 c & 0.683 & 3.19 & 0.2 & 05 37 02 & +20 43 50 & Pax \\
 & HD 37124 d & 0.6 & 1.64 & 0.14 & 05 37 02 & +20 43 50 & Concordia \\
 & HD 285968 b & 0.0265 & 0.066 & 0.0 & 04 42 56 & +18 57 29 & Cilix \\\hline
Triangulum & HD 13189 b & 14 & 1.85 & 0.28 & 02 09 40 & +32 18 59 & Acmon \\\hline
Triangulum Australe & HD 147018 b & 2.12 & 0.2388 & 0.4686 & 16 23 00 & -61 41 20 & Damnameneus \\
 & HD 147018 c & 6.56 & 1.922 & 0.133 & 16 23 00 & -61 41 20 & Celmis \\\hline
Tucana & HD 221287 b & 3.09 & 1.25 & 0.08 & 23 31 20 & -58 12 35 & Cyllene \\
 & HD 4308 b & 0.0405 & 0.118 & 0.27 & 00 44 39 & -65 38 58 & Nysa \\\hline
Ursa Major & 4 Uma b & 7.1 & 0.87 & 0.432 & 08 40 13 & +64 19 41 & Calliope \\
 & 47 Uma b & 2.6 & 2.11 & 0.049 & 10 59 29 & +40 25 46 & Clio \\
 & 47 Uma c & 0.46 & 3.39 & 0.22 & 10 59 29 & +40 25 46 & Euterpe \\
 & HAT-P-13 b & 0.851 & 0.0426 & 0.021 & 08 39 32 & +47 21 07 & Skiron \\
 & HAT-P-13 c & 15.2 & 1.186 & 0.691 & 08 39 32 & +47 21 07 & Kaikias \\
 & HAT-P-3 b & 0.599 & 0.03894 & 0.0 & 13 44 23 & +48 01 43 & Atalanta \\
 & HD 68988 b & 1.9 & 0.071 & 0.14 & 08 18 22 & +61 27 38 & Leto \\
 & HD 80606 b & 3.94 & 0.449 & 0.93366 & 09 22 37 & +50 36 13 & Niobe \\
 & HD 81688 b & 2.7 & 0.81 & 0.0 & 09 28 40 & +45 36 05 & Calydon \\
 & HD 89744 b & 7.99 & 0.89 & 0.67 & 10 22 10 & +41 13 46 & Artemis \\
 & HD 118203 b & 2.13 & 0.07 & 0.309 & 13 34 02 & +53 43 42 & Adonis \\\hline
Ursa Minor & 11 UMi b & 10.5 & 1.54 & 0.08 & 15 17 06 & +71 49 26 & Hiperborea \\
 & HD 150706 b & 1 & 0.82 & 0.38 & 16 31 17 & +79 47 23 & Boreas \\\hline
Vela & HD 73526 b & 2.9 & 0.66 & 0.19 & 08 37 16 & -41 19 08 & Argus \\
 & HD 73526 c & 2.5 & 1.05 & 0.14 & 08 37 16 & -41 19 08 & Peleus \\
 & HD 75289 b & 0.42 & 0.046 & 0.054 & 08 47 40 & -41 44 12 & Palaemon \\
 & HD 83443 b & 0.4 & 0.0406 & 0.008 & 09 37 11 & -43 16 19 & Echion \\
 & HD 85390 b & 0.14 & -- & -- & 10 00 48 & -09 31 00 & Idmon \\\hline
Virgo & 70 Vir b & 7.44 & 0.48 & 0.4 & 13 28 26 & +13 47 12 & Cora \\
 & HD 102195 b & 0.45 & 0.049 & 0.0 & 11 45 42 & +02 49 17 & Erinys \\
 & HD 102272 b & 5.9 & 0.614 & 0.05 & 11 46 24 & +14 07 26 & Demeter \\
 & HD 102272 c & 2.6 & 1.57 & 0.68 & 11 46 24 & +14 07 26 & Persephone \\
 & HD 106252 b & 6.81 & 2.61 & 0.54 & 12 13 29 & +10 02 29 & Eleusis \\
 & HD 107148 b & 0.21 & 0.269 & 0.05 & 12 19 13 & -03 19 11 & Enna \\
 & HD 110014 b & 11.09 & 2.14 & 0.462 & 12 39 14 & -07 59 44 & Callichoron \\
 & HD 114783 b & 0.99 & 1.2 & 0.1 & 13 12 43 & -02 15 54 & Nycteus \\
 & HD 130322 b & 1.08 & 0.088 & 0.048 & 14 47 32 & -00 16 53 & Alastor \\
 & HW Vir b & 19.2 & -- & -- & 12 44 20 & -08 40 17 & Hecate \\
 & HW Vir c & 8.5 & -- & -- & 12 44 20 & -08 40 17 & Rodi \\
 & PSR 1257+12 b & 7e-05 & 0.19 & 0.0 & 13 00 03 & 12 40 57 & Sisyphus \\
 & PSR 1257+12 c & 0.013 & 0.36 & 0.0186 & 13 00 03 & 12 40 57 & Ixion \\
 & PSR 1257+12 d & 0.012 & 0.46 & 0.0252 & 13 00 03 & 12 40 57 & Tantalus \\
 & WASP-16 b & 0.855 & 0.0421 & 0.0 & 14 18 44 & -20 16 32 & Liriope \\
 & HD 125612 b & 3.2 & 1.2 & 0.39 & 14 20 54 & -17 28 53 & Narcissus \\
 & HD 125612 c & 0.067 & -- & -- & 11 59 10 & -20 21 14 & Ameinias \\
 & HD 125612 d & 7.1 & -- & -- & 04 24 51 & -50 37 20 & Cephisus \\\hline
Volans & HD 76700 b & 0.197 & 0.049 & 0.13 & 08 53 55 & -66 48 03 & Nesaea \\\hline
Vulpecula & HD 188015 b & 1.26 & 1.19 & 0.15 & 19 52 04 & +28 06 01 & Laelaps \\
 & HD 189733 b & 1.13 & 0.03099 & 0.0 & 20 00 43 & +22 42 39 & Teumesia \\
 & HD 190228 b & 4.99 & 2.31 & 0.43 & 20 03 00 & +28 18 24 & Alopekos \\\hline
\end{longtable}


\vspace{1cm}

\begin{thebibliography}{}
\bibitem{}Aeschylus 467\,BC, {\it Seven Against Thebes}
\bibitem{}Alighieri, D. 1321 {\it Divine Comedy}
\bibitem{}Apollonius ca\,250\,BC, {\it Argonautica}
\bibitem{}Barlow, J. 1807 , {\it The Columbiad}  
\bibitem{}Bayer, J. 1603, {\it Uranometria}
\bibitem{}Bulfinch, T. 1855, {\it The Age of Fable, or Stories of Gods and Heroes}
\bibitem{}Cam\~oes, L. 1572, {\it Os Lus\'iadas}
\bibitem{}Cervantes, 1605, {\it Don Quixote}
\bibitem{}Cervantes, 1615, -- 
\bibitem{}Dixon-Kennedy, M. 1998, {\it Encyclopedia of Greco-Roman Mythology}, ABC-CLIO
\bibitem{}Dreyer, J. L. E. 1912. {\it The Scientific Papers of Sir William Herschel}. Royal Society and Royal Astronomical Society.
\bibitem{}Gingerish, O. 1958, ASPL,8, 9. {\it The Naming of Uranus and Neptune}
\bibitem{}Hesiod ca\,700\,BC, {\it Theogony}
\bibitem{}Hevelius, J. 1690, {\it Firmamentum Sobiescianum}
\bibitem{}Homer ca\,800\,BCa, {\it Iliad}
\bibitem{}Homer ca\,800\,BCb, {\it Odyssey}
\bibitem{}Kollestrom, N. 2009, JAHH, 12, 66. {\it The naming of Neptune}
\bibitem{}Janczak J., Fukui A., Dong S., et al., 2009, ApJ, submitted, arXiv:0908.0529 
\bibitem{}Lacaille, N., 1763, {\it Coelum Australe Stelliferum}
\bibitem{}L\'eger, A., Rouan, D., Schneider J., et al. 2009, A\&A, accepted. arXiv0908.0241
\bibitem{}L\"onnrot E. 1849, {\it The Kalevala} 
\bibitem{}Mayor, M. \& Queloz, D. 1995, Nature, 378 355
\bibitem{}Ovid 8\,AD, {\it Metamorphoses}
\bibitem{}Poincar\'e, H. 1905. {\it La valeur de la science}. Paris: Flammarion
\bibitem{}Ptolomey 148, {\it Almagest}
\bibitem{}Queloz, D., Bouchy, F., Moutou, C., et al. 2009, A\&A press release
\bibitem{}Vickers, A.M., {\it ``Joel Barlow''} in {\it Encyclopedia of American Poetry: The Nineteenth Century}. Ed. Eric Haralson. Chicago and London: Fitzroy Dearborn Publishers, 1998. 
\bibitem{}Virgil 19\,BC, {\it Aeneid}
\bibitem{}Zr\'inyi, M. 1651, {\it Siege of Sziget}
\end{thebibliography}
\end{document}